\newcommand{\beq}{\begin{equation}}
\newcommand{\eeq}{\end{equation}}
\newcommand{\beqn}{\begin{eqnarray}}
\newcommand{\eeqn}{\end{eqnarray}}
\newcommand{\Gperp}{{G_L(\sigma)}_{,x_\perp x_\perp}}
\newcommand{\Gz}{{G_L(\sigma)}_{,zz}}
\newcommand{\tGperp}{\tilde G_{L,x_\perp x_\perp}}
\newcommand{\tGz}{\tilde G_{L,zz}}
\newcommand{\Gi}{{G_L(\sigma)}_{,i}}
\newcommand{\Fperp}{F_{x_\perp x_\perp}}
\newcommand{\Fz}{F_{zz}}
\newcommand{\Fi}{F_{i}}
\newcommand{\Gperpdiv}{G_{L,x_\perp x_\perp}^{\rm div}}
\newcommand{\Gperpfin}{G_{L,x_\perp x_\perp}^{\rm fin}}
\newcommand{\Gzdiv}{G_{L,zz}^{\rm div}}
\newcommand{\Gzfin}{G_{L,zz}^{\rm fin}}
\newcommand{\Gidiv}{G_{L,i}^{\rm div}}
\newcommand{\Gifin}{G_{L,i}^{\rm fin}}

\documentstyle[aps,prd,eqsecnum,preprint]{revtex}

\begin{document}
\title{Vacuum Energy Density Fluctuations in Minkowski and Casimir States 
via Smeared Quantum Fields and Point Separation}
\author{Nicholas. G. Phillips
    \thanks{Electronic address: {\tt Nicholas.G.Phillips.1@gsfc.nasa.gov}} }
    \address{Raytheon ITSS, Laboratory for Astronomy and Solar Physics, Code 685,
             NASA/GSFC, Greenbelt, Maryland 20771}
\author{B. L. Hu 
    \thanks{Electronic address: {\tt hub@physics.umd.edu}} }
    \address{Department of Physics, University of Maryland, College Park, Maryland 20742-4111}
\date{submitted to Phys. Rev. D, May 31, 2000}
\maketitle

\begin{abstract}
We present calculations of the variance of fluctuations and of the mean of the
energy  momentum tensor of a massless scalar field  for  the Minkowski and
Casimir vacua as a function of an intrinsic  scale defined by  a smeared field
or by point separation. We point out that contrary to prior claims,
 the ratio of variance to mean-squared being
of the order unity is not necessarily a good criterion for measuring the invalidity of
semiclassical gravity. For the Casimir topology we obtain expressions for the
variance  to mean-squared ratio  as a function of the intrinsic scale (defined by a
smeared field)  compared to the extrinsic scale (defined by the separation of
the plates, or the periodicity of space).
Our results make it possible to identify
the spatial extent where negative energy density prevails which could be useful
for studying  quantum field effects in worm holes and baby universe,
and for examining the design feasibility of  real-life `time-machines'.
 For the Minkowski vacuum we find
that  the ratio of the variance to the mean-squared, calculated from the
coincidence limit, is identical to the value of the Casimir case at the same
limit for spatial point separation  while identical to the value of a hot flat
space result with a temporal point-separation. We analyze the origin of
divergences in the fluctuations of the energy density and discuss choices in
formulating a procedure for their removal,  thus raising new questions into the
uniqueness and even the very meaning of  regularization of the energy momentum
tensor  for quantum fields in curved or even flat spacetimes when spacetime is
viewed as having  an extended structure.
\end{abstract}

\section{Introduction}
Recent years saw the beginning of serious studies of the fluctuations of the
energy momentum tensor (EMT) $\hat T_{\mu \nu}$
of quantum fields in spacetimes with boundaries \cite{qftcst}
(such as Casimir effect \cite{Casimir}) \cite{Barton,KuoFor}, nontrivial
topology (such as imaginary time thermal field theory) or nonzero curvature
(such as the Einstein universe) \cite{PH97}. 
A natural measure of the strength of fluctuations is $\chi$ \cite{HP0},
 the ratio of the variance $\Delta \rho^2$
of  fluctuations in the energy density (expectation value of the $\hat \rho^2$ 
operator minus the square
of the mean $\hat \rho$ taken with respect to some quantum state) to  its 
mean-squared (square of the 
expectation value of $\hat \rho$):
\beq
\chi \equiv  \frac{\left<\hat\rho^2\right>-\left<\hat\rho\right>^2}
               {\left<\hat\rho\right>^2} 
       \equiv \frac { \Delta \rho^2} {{\bar \rho}^2}
\eeq
Alternatively, we can use the quantity introduced by Kuo and Ford \cite{KuoFor}  
\begin{equation}
\Delta \equiv \frac{\left<\hat\rho^2\right>-\left<\hat\rho\right>^2}
               {\left<\hat\rho^2\right>} = \frac {\chi} {\chi +1}
\end{equation}
Assuming a positive definite variance $\Delta \rho^2 \ge 0$, 
then $ 0 \le \chi \le \infty$  and $0 \le\Delta \le 1$ always, 
with $\Delta \ll 1$  falling in the classical domain. 
Kuo and Ford (KF)  displayed  a number of quantum states (vacuum plus 2 particle
state, squeezed vacuum and Casimir vacuum) with respect to which the
expectation value of the energy momentum tensor (00 component) gives rise to
negative local energy density. For these states  $\Delta$ is of order unity. 
From this result they drew the implications, amongst other interesting
inferences, that semiclassical gravity (SCG) \cite{scg} based on the semiclassical
Einstein equation 
\begin{equation}
G_{\mu\nu} = 8\pi G \left< \hat T_{\mu\nu} \right>
\end{equation}
(where $G_{\mu \nu}$ is the Einstein tensor and $G$ the Newton gravitational constant)
could become invalid under these conditions. The validity of semiclassical
gravity in the face of fluctuations of quantum fields as source is an important
issue which has caught the attention of many authors \cite{valSCG}. We hold a
different viewpoint on this issue from KF, which we hope to clarify in this
study.

In this series of papers, we would like to examine more closely  this and
related issues of SCG,  such as the regularization in the  energy momentum
tensor of quantum fields and fluctuations of spacetime metric.
In this paper we study a free field in flat space and  spacetimes with
boundaries. Later papers deal with curved spacetimes depicting the early universe
(quantum fluctuations and structure formation) and black holes (horizon
fluctuations and dynamical backreactions).

As explained elsewhere by one of us, when fluctuations of the energy momentum
tensors are included as source for the dynamics of spacetime, these problems
are best discussed in the larger context of stochastic semiclassical gravity
(SSG) \cite{stogra} program based on Einstein-Langevin type of equations
\cite{ELE}, which is the proper framework to address Planck scale physics.

There are two groups of interrelated issues in quantum field theory in flat
(ordinary QFT) or curved spacetimes (QFTCST), or semiclassical gravity (SCG --where
the background spacetime dynamics is determined by the backreaction  of the
mean value of quantum fields): one pertaining to quantum fields and the other
to spacetimes. We discuss the first set relating to the fluctuations of the EMT
over its mean values with respect to the vacuum state. It strikes us as no
great surprise that  
states which are more quantum (e.g., squeezed states) in nature  than
classical (e.g., coherent states) \cite{states} may lead to large 
fluctuations comparable to the mean in the
energy density. This can be seen even in the ratio of expectation values of
moments of the displacement operators in simple  quantum harmonic oscillators
\cite{Raval}. Such a condition exists peacefully with the underlying spacetime
at least at the low energy of today's universe. We don't see sufficient ground
to question the validity of SCG at energy below the Planck energy when the
spacetime is depictable  by a manifold structure,  approximated locally by
the Minkowski space. Besides, the cases studied in Kuo and Ford \cite{KuoFor}
as well as many others \cite{PH97,valSCG} are of a test-field nature, where
backreaction is not  considered. (So KF's criterion pertains more to QFTCST
than to SCG, where in the former the central issue is compatibility, which is a
weaker condition than consistency in the latter.)

To assess  this situation we aim at calculating the variance of fluctuations
to mean-squared ratio of a quantum field for the simplest case of Minkowski spacetime 
i.e., for ordinary quantum field theory. We find that $\Delta=2/5$. 
This is a clear-cut counter-example to the claim of KF,
since $\Delta = O(1)$ holds also for Minkowski  space, where SCG is known to be
valid at large scales.  We view this situation as arising from  the quantum nature of 
the vacuum state and saying little about the compatibility of the field source
with the spacetime the quantum field lives in. In contrast, our view on this
issue is that one should refer to a scale (of interaction or for probing
accuracy) when measuring the validity of SCG. The conventional belief is that
when reaching the Planck scale from below, QFTCST will break down because,
amongst other things happening,  graviton production at that energy will become
significant so as to render the classical background spacetime unstable, and
the mean value of quantum field taken as a source for the Einstein equation
becomes inadequate. 
To address this issues as well as the issue of the spatial extent
where negative energy density can exist,  we view it necessary to introduce a scale in the
spacetime regions where quantum fields are defined to monitor how the mean value
and the fluctuations of the energy momentum tensor change. Point separation
is an ideal method to adopt for this purpose. 

In conventional field theories 
the stress tensor  built from the product of a pair of field operators
evaluated at a single point in the spacetime manifold  is, strictly speaking, 
ill-defined. The point separation scheme \cite{ptsep} was
introduced as a method of regularization of the energy momentum tensor for
quantum fields in curved spacetime.  In this scheme, one introduces an
artificial separation of the single point $x$ to a pair of closely separated
points $x$ and $x'$. The problematic terms involving field products such as
$\hat\phi(x)^2$ becomes $\hat\phi(x)\hat\phi(x')$, whose expectation value is
well defined. One then brings the two points back (taking the coincidence
limit) to identify the divergences present, which will then be removed
(regularization) or moved (by renormalizing the coupling constants), thereby
obtaining  a well-defined, finite stress tensor at a single point. In this
context point separation was introduced as a trick for  identifying the
ultraviolet divergences in a covariant manner. One of us in the development
of  the stochastic semiclassical gravity program \cite{stogra} has maintained 
the view (and  as we will expound further in later papers ) that  instead of
being used as a mere technical device in  QFTCST, this method has  much greater
physical content. We prefer to view the operator valued  EM bi-tensor $\hat
T_{\mu\nu} (x,y)$ and EM two point function  $\left< \hat T_{\mu\nu}
(x,y)\right>$ as the  more fundamental objects in a more basic theory of
spacetime and matter which has the point-defined quantum field theory as a low
energy limit. This allows for spacetime to acquire an extended structure at
sub-Planckian scale.\footnote{Viewing the two points here as the end points of an open 
string gives what one of us called a `skeletal representation' of string theory, 
where the internal excitation modes are suppressed. Related to the so-called 
dipole approximation of string theory in recent literature \cite{dipolestring}
it should also carry features of non-commutative
geometry of spacetime possible at the Planck scale.}

For our stated purpose above, there is another way to  introduce a scale in the
quantum field theory, i.e.,
by introducing a (spatial) smearing  function $f({\bf x}) $ to define smeared field
operators $\hat \phi_t(f_{\bf x})$. In this paper we shall construct a scheme to encompass
both aspects, by defining field operators at two separated points (connected by
distance $r$) and using a Gaussian smearing  function (with variance
$\sigma^2$). We derive expressions for the EM bi-tensor operator, its mean and
its fluctuations as  functions of $r,\sigma$, for a massless scalar field in
both the Minkowski and the Casimir spacetimes.  The interesting result we find 
is that while both the vacuum expectation value and the  fluctuations of energy
density grow as $\sigma \rightarrow 0$,  the ratio of the variance of the
fluctuations to its mean-squared remains a constant $\chi_d$ ($d$ is the spatial
dimension of spacetime) which is independent of $\sigma$. The measure $\Delta_d$
($=\chi_d/(\chi_d+1)$) depends on the dimension of space and is of the order unity.
It varies only slightly for spacetimes with boundary or nontrivial topology. 
For example $\Delta$ for Minkowski is $2/5$, while for Casimir is $6/7$
(cf,  from \cite{PH97}). Add to this our prior result for the Einstein Universe, 
 $ \Delta=111/112$, independent of curvature, and that for hot flat space
\cite{NPhD}, we see a pattern emerging.

These results allow us to address two  interrelated issues:
1) Fluctuations of the energy density and validity of semiclassical gravity,
and  2) The spatial extent where negative energy density can exist. For Issue 1)
we see that i) the fluctuations of the energy density as well as its mean both 
increase with decreasing distance (or probing scale), while ii)
the ratio of the variance of the fluctuations in EMT to its mean-squared 
is of the order unity. 
We view the first but not the second feature as linked to the question of the validity of SCG
--the case for Minkowski spacetime alone is sufficient to testify to the fallacy of
Kuo and Ford's criterion. The second feature represents something quite different,  
pertaining more to the quantum nature of the  vacuum state than to the validity of SCG.

For Issue 2) it is well known that negative energy density exists in Casimir geometry,
moving mirrors, black holes and worm holes. Proposals have also been conjured
to use the negative energy density for the design of time machines \cite{Thorne}. Our results
(Figures 1, 2) provide an explicit scale dependence of the regularized vacuum energy density 
$\rho_{L,reg}$ and its fluctuations $\Delta_{L,reg}$ , specifically $\sigma/L$, the ratio
of the smearing length (field scale) to that of the Casimir length (geometry scale).  For 
example, Fig. 2 shows that only for $\sigma /L < 0.24$ is $\rho_{L,reg}<0$.
Recall $\sigma$ gives the spatial extent the field is probed or smeared. Ordinary
pointwise quantum field theory which probes the field only at a point does not carry
information about the spatial extent where negative energy density sustains.
These results have direct implications on wormhole physics (and time machines,  if one 
gets really serious about these fictions \cite{Thorne}). 
If L is the scale characterizing the size (`throat') of the wormhole where one
thinks negative energy density might prevail, and designers of `time machines'
wish to exploit for `time-travel' , our result provides a limit
on the size of the probe (spaceship in the case of time-travel) in ratio to L where such conditions
may exist. It could also provide a quantum field-theoretical bound on the probability
of spontaneous creation of baby universes from quantum field energy fluctuations. 

An equally weighty issue brought to light in this study is  3) the meaning of
regularization in the face of EMT fluctuations. Since we have the
point-separated expressions of the EMT and its fluctuations we can study how
they change as a function of separation or smearing. In particular we can see
how  divergences arise at the coincidence limit. Whether certain cross terms
containing divergences have physical meaning is a question raised by the recent
studies of Wu and Ford \cite{WuFor}.  We can use these calculations to examine
these issues and  ask the broader question of what exactly  regularization
entails, where divergences arise and how they are to be treated. The 
consideration  of divergences in the  fluctuations of EMT  requires  a more
sophisticated rationale and reveals a deeper  layer of issues pertaining to
effective versus more fundamental theories. If we view ordinary quantum field
theory defined at points as a  low energy limit of a  theory of spacetime
involving extended structures (such as string theory), then these results would
shed light on their meaning and inter-connections.

In this paper we will discuss  three aspects of quantum field theory 
in curved spacetime in the light of fluctuations of quantum stress energy :
1) Fluctuation to mean ratio of vacuum energy density  and the validity of SCG;
2) The point separated results of the mean and the fluctuations
of the energy density for two states: the Minkowski case which has
no scale present (massless field) and the Casimir case which 
has a scale present (the separation of the plates);
3) The circumstances when and how divergences appear and the 
meaning of regularization in point-defined field theories versus theories 
defined at separated points and/or smeard fields.
A summary of the main points of 1) has been given in \cite{HP0}.
In this paper we give details of the calculations  and discuss the
regularization issue 3), while leaving the issue 2) of the spatial extent of
negative energy density and its implications for quantum effects
of worm holes, baby universes and time travel to a future investigation.
In Sec. 2 we define the smeared field operators and their products defined at
separated points. In Sec. 3  we construct from these the smeared energy density 
and its fluctuations,
and calculate the ratio of the fluctuations to the mean for a flat space (Minkowski
geometry). 
In Sec. 4 we analyze the case for  a Casimir geometry of one periodic spatial dimension.
In Sec. 5 we consider the point-separation calculation in  Minkowski  space,
comparing the different results obtained from taking the coincidence limit from temporal versus 
spatial directions. Finally in Sec. 6 we summarize the major results and discuss the meaning of 
our finding in relation to the issues raised above.

\section{Smeared Field Operators at Separated Points}

Since the field operator in conventional point-defined quantum field theory is an
operator-valued distribution, products of field operators
at a point become problematic. This parallels the problem with
defining the square of a delta function $\delta^2(x)$. Distributions
are defined via their integral against a test function: they live
in the space dual to the test function space. By going from the
field operator $\hat\phi(x)$ to its integral against a test function,
$\hat\phi(f) = \int \hat\phi\,f$, we can now readily consider products.

When we take the test functions to be spatial Gaussians, we are
smearing the field operator over a finite spatial region.
Physically we 
see  smearing as representing the necessarily finite extent of an 
observer's probe, or the intrinsic limit of resolution in carrying out 
a measurement at a low energy (compared to Planck scale).
In contrast to the ordinary  point-defined quantum field theory,
where ultraviolet divergences occur in the energy momentum tensor, 
smeared fields give no ultraviolet divergence. This is because smearing is 
equivalent to a regularization scheme which imparts an exponential 
suppression to the high momentum modes  and restricts the contribution 
of the high  frequency modes in the mode sum.

With this in mind, we start by defining the spatially smeared field operator
\beq
\hat\phi_t(f_{\bf x}) = 
     \int \hat\phi(t,{\bf x'}) f_{\bf x}({\bf x}')  d{\bf x}'
\eeq
where $f_{\bf x}({\bf x}')$ is a suitably smooth function.
With this, the two point operator becomes
\beq
\left(\hat\phi_t(f_{\bf x})\right)^2 = \int\int 
 \hat\phi(t,{\bf x}') \hat\phi(t,{\bf x}'') 
        f_{\bf x}({\bf x}') f_{\bf x}({\bf x}'')
  d{\bf x}\,d{\bf x}'
\eeq
which is now finite.
In terms of the vacuum $\left|\left.0\right>\right.$
($\hat a_{\bf k}\left|\left.0\right>\right. = 0 $, for all ${\bf k}$)
we have the usual mode expansion
\beq
\hat \phi\left(t_1,{\bf x}_1\right) 
= \int d\mu\left({\bf k}_1\right) \left(
 \hat a_{{\bf k}_{1}}\,u_{{\bf k}_{1}}\!\left(t_1,{\bf x}_1\right) + 
  \hat a^{\dagger}_{{\bf k}_{1}}\,u^{*}_{{\bf k}_{1}}\!\left(t_1,{\bf x}_1\right)
\right)
\eeq
with
\beq
  u_{{\bf k}_{1}}\left(t_1,{\bf x}_1\right) = 
	N_{k_{1}} {e^{i\,\left( {\bf k}_{1}\cdot{\bf x}_{1} -
                       t_{1}\,\omega_{1} \right) }}, 
\quad
  \omega_1 = \left| {\bf k}_1 \right|,
\eeq
where the integration measure $\int d \mu\left({\bf k}_1\right)$ and the 
normalization constants $N_{{\bf k}_1}$ are given for a Minkowski
and Casimir spaces by (\ref{Mdu}) and  (\ref{Cdu}) respectively.

In this work, we use a Gaussian smearing function
\beq
f_{{\bf x}_0}({\bf x}) 
 = \left(
{\frac{1}{4\,\pi \,{{\sigma}^2}}} \right)^\frac{d}{2}
{e^{- \left(\frac{{\bf x}_{0} - {\bf x}_{1}}{2\sigma}\right)^2}}
\eeq
with the properties
$\int f_{{\bf x}_0}\left({\bf x}'\right) d{\bf x}' = 
1$,
$\int {\bf x}'\,f_{{\bf x}_0}\left({\bf x}'\right) 
d{\bf x}' = {\bf x}_0$ and
$\int |{\bf x}'|^2\,f_{{\bf x}_0}\left({\bf x}'\right) 
d{\bf x}' =  2d\sigma^2 + |{\bf x}_0|^2$.
Using
\beqn
\int 
  u_{{\bf k}_{1}}\!\left(t,{\bf x}\right) 
    f_{{\bf x}_1}({\bf x})
d{\bf x} 
&=&
  N_{k_{1}}  e^{-it\omega_1}\;
  \prod_{i=1}^d\left( 
     {\frac{1}{2\,{\sqrt{\pi }}\,\sigma}}
    \int e^{+ik_{1i} x_i - 
                     \left(\frac{x_{1i}-x_{i}}{2\sigma}\right)^2
           } \,dx_{i} 
  \right) 
\nonumber \\
&=&
N_{k_{1}}
{e^{-i\,t\,w + i\,{\bf k}_{1}\cdot{\bf x}_{1} - {{\sigma}^2}\,{{k_{1}}^2}}}
\eeqn
we get the smeared field operator
\beq
\hat\phi_{t_1}\left(f_{{\bf x}_1}\right) = \int d\mu\left({\bf k}_1\right) 
N_{k_{1}}
{e^{-i\,{\bf k}_{1}\cdot{\bf x}_{1} - {{\sigma}^2}\,{{k_{1}}^2} - 
      i\,t_{1}\,\omega_{1}}}\,\left( {e^{2\,i\,{\bf k}_{1}\cdot{\bf x}_{1}}}\,
     \hat a_{{\bf k}_{1}} + {e^{2\,i\,t_{1}\,\omega_{1}}}\,
     \hat a^{\dagger}_{{\bf k}_{1}} \right) 
\eeq
and their  product:
\beqn
\hat\phi_{t_1}\left(f_{{\bf x}_1}\right)
\hat\phi_{t_2}\left(f_{{\bf x}_2}\right)
&=& \int d\mu\left({\bf k}_1,{\bf k}_2\right) N_{k_{1}}\,N_{k_{2}}
{e^{-{{\sigma}^2}\,\left( {{k_{1}}^2} + {{k_{2}}^2} \right)
         - i\,\left( {\bf k}_{1}\cdot{\bf x}_{1} + 
        {\bf k}_{2}\cdot{\bf x}_{2} + t_{1}\,\omega_{1} + t_{2}\,\omega_{2}
         \right) }} \nonumber \\
&& \left( 
   {e^{2\,i\,\left( {\bf k}_{1}\cdot{\bf x}_{1} + 
         {\bf k}_{2}\cdot{\bf x}_{2} \right) }}\,
   \hat a_{{\bf k}_{1}}\hat a_{{\bf k}_{2}} + 
  {e^{2\,i\,\left( {\bf k}_{1}\cdot{\bf x}_{1} + t_{2}\,\omega_{2} \right)
         }}\,\hat a_{{\bf k}_{1}}\hat a^{\dagger}_{{\bf k}_{2}} + 
  {e^{2\,i\,\left( {\bf k}_{2}\cdot{\bf x}_{2} + t_{1}\,\omega_{1} \right)
         }}\,\hat a^{\dagger}_{{\bf k}_{1}}\hat a_{{\bf k}_{2}} 
\right. \nonumber \\ && \left.
 + {e^{2\,i\,\left( t_{1}\,\omega_{1} + t_{2}\,\omega_{2} \right) }}\,
  \hat a^{\dagger}_{{\bf k}_{1}}\hat a^{\dagger}_{{\bf k}_{2}} 
\right).
\eeqn
The smearing has introduced a factor $e^{-\sigma^2\,k^2}$ for each of the momenta.
This factor acts as the regulator for the mode sums; in this sense the high
momenta that led to the divergences have been controlled. We also note the
$\sigma\rightarrow 0$ limit amounts to the relaxation of the regulator and
the point field theory version of the field operator is recovered.

\subsection{Coincident Smeared Green Function}

We can let the two spatial points come together, ${\bf x}_2 = {\bf x}_1 = 0$,
and get the coincident limit of the Green function
\beqn
G\left(\Delta t \right) &=&
\left<0\left| 
  \hat\phi_{t_1}\left(f_{{\bf x}_1}\right)
  \hat\phi_{t_2}\left(f_{{\bf x}_1}\right)
\right|0\right> \nonumber \\
&=& \int d\mu\left({\bf k}_1\right) {{N^2_{k_{1}}}}
{e^{-2\,{{\sigma}^2}\,{{k^2_{1}}} + i\,{\Delta t}\,\omega_{1}}}
\eeqn
For flat space the normalization and integration measure are
\beq
{{N^2_{k_{1}}}} = {\frac{1}{{2^{d+1}{\pi }^d}\,\omega_{1}}}
\quad{\rm and}\quad
\int d\mu({\bf k}_1) = {\frac{2\,{{\pi 
}^{{\frac{d}{2}}}}}{\Gamma\!\left({\frac{d}{2}}\right)}} \int_0^\infty k_1^{d-1} 
dk_1
\label{Mdu}
\eeq
and the Green function is
\begin{mathletters}
\beqn
G({\Delta t}) &=&
\frac{1}{{2^{{\frac{3}{2}\,\left( d+1 \right)}}}\,{{\pi }^{{\frac{d}{2}}}}\,
  {{\sigma}^d}\,\Gamma\!\left({\frac{d}{2}}\right)}
\left\{
  2\,\sigma\,\Gamma\!\left({\frac{d-1}{2}}\right)\,
  {}_{1}F_{1}\left({\frac{d-1}{2}};{\frac{1}{2}};
   -{\frac{{{{\Delta t}}^2}}{8\,{{\sigma}^2}}}\right)
\right. \nonumber \\ && \hspace{35mm} \left.
 + i\,{\sqrt{2}}\,{\Delta t}\,\Gamma\!\left({\frac{d}{2}}\right)\,
  {}_{1}F_{1}\left({\frac{d}{2}};{\frac{3}{2}};
   -{\frac{{{{\Delta t}}^2}}{8\,{{\sigma}^2}}}\right)
\right\} \\ \nonumber \\
&=&
{\frac{
    \Gamma\!\left({\frac{d-1}{2}}\right)
      }{{2^{{\frac{3d+1}{2}}}}\,
      {\pi }^{{\frac{d}{2}}}{{\sigma}^{d - 1}}\,\,\Gamma\!\left({\frac{d}{2}}\right)}
      }
\left( 1
+ {\frac{i\,{\Delta t}\,\Gamma\!\left({\frac{d}{2}}\right)}
   {{\sqrt{2}}\,\sigma\,\Gamma\!\left({\frac{d-1}{2}}\right)}}
- {\frac{\left( d-1 \right) \,{{{\Delta t}}^2}}{8\,{{\sigma}^2}}}
\right) + O\left( \Delta t^3\right),
\eeqn
\end{mathletters}
which we see is finite for this spatial coincident limit, and is
finite for the $\Delta t\rightarrow 0$ limit as well.
For spatial dimension $d=3$, the smeared Green function is
\begin{mathletters}
\beqn
G({\Delta t}) &=&
{\frac{1}{16\,{{\pi }^2}\,{{\sigma}^2}}}
- {\frac{{\Delta t}\,\left({\rm Erfi}({\frac{{\Delta t}}
          {2\,{\sqrt{2}}\,\sigma}}) -i\right) }{32\,{\sqrt{2}}\,
     {e^{{\frac{{{{\Delta t}}^2}}{8\,{{\sigma}^2}}}}}\,
     {{\pi }^{{\frac{3}{2}}}}\,{{\sigma}^3}}} \\ \nonumber \\
&=&
{\frac{1}{16\,{{\pi }^2}\,{{\sigma}^2}}}
\left( 1
+ {\frac{i{\Delta t}\,{\sqrt{\pi}}}{2\sqrt{2}\sigma}}
- {\frac{{{{\Delta t}}^2}}{4\,{{\sigma}^2}}}
\right) + O\left( \Delta t^3\right)
\eeqn
\end{mathletters}

\subsection{Point-Separated Smeared Energy Density Operator}

For a classical scalar function, the energy density is
\beq
\rho\left(t_1,{\bf x}_1\right) = \frac{1}{2}\left(
 \left( \partial_{t_1} \phi \right)^2
+ \left( \vec\nabla \phi \right)^2
\right)
\eeq
We cannot go directly to the quantum field case since the energy density
has pairs of field operators evaluated at the same point. We can however define
an energy density operator which contains smeared field operators at 
separated points. Then by taking the coincident (${\bf x}_1 \rightarrow {\bf x}_2$)
limit, we will be using the smearing to regularize the energy density, while
if we take the zero smearing width  ($\sigma\rightarrow 0$) limit,
we are using point separation regularization.

Point separation consists of symmetrically splitting the operator product as
\beq
\hat\phi(t_1,{\bf x}_1)^2 \rightarrow \frac{1}{2}
 \left(
  \hat\phi(t_1,{\bf x}_1)\hat\phi(t_2,{\bf x}_2)
              + \hat\phi(t_2,{\bf x}_2)\hat\phi(t_1,{\bf x}_1)
 \right).
\eeq
Products of derivatives of the field operator are symmetrically split according
to, e.g., time derivatives become
\beq
\left( \partial_{t_1} \hat\phi(t_1,{\bf x}_1) \right)^2 \rightarrow \frac{1}{2}
 \left(
 \left( \partial_{t_1}\hat\phi(t_1,{\bf x}_1) \right)
 \left( \partial_{t_2}\hat\phi(t_2,{\bf x}_2) \right)
+\left( \partial_{t_2}\hat\phi(t_2,{\bf x}_2) \right)
 \left( \partial_{t_1}\hat\phi(t_1,{\bf x}_1) \right)
 \right)
\eeq
We introduce the smeared field operator derivatives
\begin{mathletters}
\beqn
\left(\partial_{t_1}\hat\phi_{t_1}\right)\left(f_{{\bf x}_1}\right) &=& \int 
   \left(\partial_{t_1}\hat\phi\left(t_1,{\bf x}'\right)\right)
   f_{{\bf x}_1}\left({\bf x}'\right)
\, d{\bf x}'
\nonumber \\
&=& i \int d\mu({\bf k}_1)
N_{k_{1}}\,\omega_{1}
{e^{-i\,{\bf k}_{1}\cdot{\bf x}_{1} - {{\sigma}^2}\,{{k_{1}}^2} - 
     i\,t_{1}\,\omega_{1}}} \left( {e^{2\,i\,t_{1}\,\omega_{1}}}\,\hat 
a^{\dagger}_{{\bf k}_{1}} - {e^{2\,i\,{\bf k}_{1}\cdot{\bf x}_{1}}}\,\hat 
a_{{\bf k}_{1}} \right)
\\ \nonumber \\
\left(\vec\nabla_{{\bf x}_1}\hat\phi_{t_1}\right)\left(f_{{\bf x}_1}\right) &=& \int 
   \left(\vec\nabla_{{\bf x}'}\hat\phi\left(t_1,{\bf x}'\right)\right)
   f_{{\bf x}_1}\left({\bf x}'\right)
\, d{\bf x}'
\nonumber \\
&=& -i \int d\mu({\bf k}_1)
{\bf k}_{1}\,N_{k_{1}}
{e^{-i\,{\bf k}_{1}\cdot{\bf x}_{1} - {{\sigma}^2}\,{{k_{1}}^2} - 
     i\,t_{1}\,\omega_{1}}} \left( {e^{2\,i\,t_{1}\,\omega_{1}}}\,\hat 
a^{\dagger}_{{\bf k}_{1}} - {e^{2\,i\,{\bf k}_{1}\cdot{\bf x}_{1}}}\,\hat 
a_{{\bf k}_{1}} \right)
\eeqn
\end{mathletters}
and use the symmetric splitting to
define the point separated smeared energy density operator
\beqn
\hat\rho\left(t_1,{\bf x}_1; t_2,{\bf x}_2;\sigma\right) &=& 
 \frac{1}{4}\left\{
    \left(\left(\partial_{t_1}\hat\phi_{t_1}\right)\left(f_{{\bf x}_1}\right)\right) 
    \left(\left(\partial_{t_2}\hat\phi_{t_2}\right)\left(f_{{\bf x}_2}\right)\right) 
+
    \left(\left(\partial_{t_2}\hat\phi_{t_2}\right)\left(f_{{\bf x}_2}\right)\right) 
    \left(\left(\partial_{t_1}\hat\phi_{t_1}\right)\left(f_{{\bf x}_1}\right)\right)
\right. \nonumber \\ &&\hspace{5mm}\left.
+
  \left(\left(\vec\nabla_{{\bf x}_1}\hat\phi_{t_1}\right)\left(f_{{\bf x}_1}\right)\right) 
  \left(\left(\vec\nabla_{{\bf x}_2}\hat\phi_{t_2}\right)\left(f_{{\bf x}_2}\right)\right) 
+
  \left(\left(\vec\nabla_{{\bf x}_2}\hat\phi_{t_2}\right)\left(f_{{\bf x}_2}\right)\right) 
  \left(\left(\vec\nabla_{{\bf x}_1}\hat\phi_{t_1}\right)\left(f_{{\bf x}_1}\right)\right) 
\right\}
\nonumber \\
&=&
 -{\frac{1}{4}} \int d\mu({\bf k}_1)\,d\mu({\bf k}_2)
N_{k_{1}}\,N_{k_{2}}\,\left( {\bf k}_{1}\cdot{\bf k}_{2} + 
    \omega_{1}\,\omega_{2} \right) 
\nonumber \\
&&\times
{e^{
   -i\left({\bf k}_{1}\cdot{\bf x}_{1} + {\bf k}_{2}\cdot{\bf x}_{2}\right) 
   - i\left(t_{1}\,\omega_{1} + t_{2}\,\omega_{2}\right)
   - {{\sigma}^2}\left({{k_{1}}^2} + {{k_{2}}^2}\right) 
   }}
\nonumber \\
&&\times\left(
  {e^{2\,i\,{\bf k}_{1}\cdot{\bf x}_{1} + 2\,i\,{\bf k}_{2}\cdot{\bf x}_{2}}}\,
   \hat a_{{\bf k}_{1}}\hat a_{{\bf k}_{2}} - 
  {e^{2\,i\,{\bf k}_{1}\cdot{\bf x}_{1} + 2\,i\,t_{2}\,\omega_{2}}}\,
   \hat a_{{\bf k}_{1}}\hat a^{\dagger}_{{\bf k}_{2}}
\right.\nonumber \\&&\hspace{3mm}
+ {e^{2\,i\,{\bf k}_{1}\cdot{\bf x}_{1} + 2\,i\,{\bf k}_{2}\cdot{\bf x}_{2}}}\,
   \hat a_{{\bf k}_{2}}\hat a_{{\bf k}_{1}} - 
  {e^{2\,i\,{\bf k}_{2}\cdot{\bf x}_{2} + 2\,i\,t_{1}\,\omega_{1}}}\,
   \hat a_{{\bf k}_{2}}\hat a^{\dagger}_{{\bf k}_{1}}
\nonumber \\&&\hspace{3mm}
  - {e^{2\,i\,{\bf k}_{2}\cdot{\bf x}_{2} + 2\,i\,t_{1}\,\omega_{1}}}\,
     \hat a^{\dagger}_{{\bf k}_{1}}\hat a_{{\bf k}_{2}}   + 
  {e^{2\,i\,t_{1}\,\omega_{1} + 2\,i\,t_{2}\,\omega_{2}}}\,
   \hat a^{\dagger}_{{\bf k}_{1}}\hat a^{\dagger}_{{\bf k}_{2}}
\nonumber \\&&\hspace{3mm}\left.
  - {e^{2\,i\,{\bf k}_{1}\cdot{\bf x}_{1} + 2\,i\,t_{2}\,\omega_{2}}}\,
     \hat a^{\dagger}_{{\bf k}_{2}}\hat a_{{\bf k}_{1}}   + 
  {e^{2\,i\,t_{1}\,\omega_{1} + 2\,i\,t_{2}\,\omega_{2}}}\,
   \hat a^{\dagger}_{{\bf k}_{2}}\hat a^{\dagger}_{{\bf k}_{1}}
\right)
\eeqn
Its vacuum expectation value is
\beqn
\rho\left(t_1,{\bf x}_1; t_2,{\bf x}_2;\sigma\right) &=&
     \left<0\left|
         \hat\rho_t\left(t_1,{\bf x}_1; t_2,{\bf x}_2;\sigma\right) 
     \right|0\right>
\nonumber \\
&=& {\frac{1}{2}}\int d\mu\left({\bf k}_1\right) 
{{N^2_{k_{1}}}}\,{{\omega^2_{1}}}\,
{e^{-2\,{{\sigma}^2}\,{{k^2_{1}}}}}
\nonumber \\&&\hspace{5mm}\times
\left( {e^{-i\,\left( {\bf k}_{1}\cdot\left({\bf x}_{1} - {\bf x}_{2}\right) - 
        \left( t_{1} - t_{2} \right) \,\omega_{1} \right) }} + 
  {e^{i\,\left( {\bf k}_{1}\cdot\left({\bf x}_{1} - {\bf x}_{2}\right) -
        \left( t_{1} - t_{2} \right) \,\omega_{1} \right) }} \right)
\label{energy-density-general}
\eeqn
We obtain the smeared vacuum energy density by taking the coincidence
\beq
\rho\left(\sigma\right) =
\int d\mu\left({\bf k}_1\right) {{N^2_{k_{1}}}}\,{{\omega^2_{1}}} \,
{e^{-2\,{{\sigma}^2}\,{{k^2_{1}}}}}
\label{rho-sigma}
\eeq
while the point separation expression is obtained from taking the zero 
smearing-width limit:
\beq
\rho\left(t_1,{\bf x}_1; t_2,{\bf x}_2\right) =
\int d \mu ({\bf k}_1)  {{N^2_{k_{1}}}}\,{{\omega^2_{1}}} \,
  \cos ({\bf k}_{1}\cdot\left({\bf x}_{1} - {\bf x}_{2}\right) -
   \left(t_{1} - t_{2} \right) \,\omega_{1})
\label{rho-xx}
\eeq

\subsection{Point-Separated Smeared Energy Density Correlation Function}

We now consider the point separated vacuum correlation function for the
energy density operator:
\beqn
\Delta\rho^2(t_1,{\bf x}_1,t'_1,{\bf x}'_1;t_2,{\bf x}_2,t'_2,{\bf x}'_2;\sigma)
&=&
\left<0\left|
\hat\rho\left(t_1,{\bf x}_1; t'_1,{\bf x}'_1;\sigma\right)
\hat\rho\left(t_2,{\bf x}_2; t'_2,{\bf x}'_2;\sigma\right)
\right|0\right>
\nonumber \\
&&
- \rho\left(t_1,{\bf x}_1; t'_1,{\bf x}'_1;\sigma\right)
  \rho\left(t_2,{\bf x}_2; t'_2,{\bf x}'_2;\sigma\right)
\eeqn
With this definition, the vacuum energy density correlation function 
is
\beqn
\Delta\rho^2(t_1,{\bf x}_1;t_2,{\bf x}_2)
  &\equiv&
    \left<0\left|
      \hat\rho\left(t_1,{\bf x}_1\right)\hat\rho\left(t_2,{\bf x}_2\right)
    \right|0\right>    
-
    \left<0\left|
      \hat\rho\left(t_1,{\bf x}_1\right)
    \right|0\right>    
    \left<0\left|
      \hat\rho\left(t_2,{\bf x}_2\right)
    \right|0\right>
\nonumber \\
&=& 
   \Delta\rho^2\left(
        t_1, {\bf x}_1, t_1, {\bf x}_1;
        t_2, {\bf x}_2, t_2, {\bf x}_2;
        \sigma = 0
   \right)
\eeqn
Since the divergences present in
$    \left<0\left|
      \hat\rho\left(t_1,{\bf x}_1\right)\hat\rho\left(t_2,{\bf x}_2\right)
    \right|0\right>$
for $\left(t_2,{\bf x}_2\right) \ne \left(t_1,{\bf x}_1\right)$
are canceled by those due to
$\left<0\left|\hat\rho\left(t_1,{\bf x}_1\right)\right|0\right>$ and
$\left<0\left|\hat\rho\left(t_2,{\bf x}_2\right)\right|0\right>$, we can
assume $\left(t'_1,{\bf x}'_1\right) = \left(t_1,{\bf x}_1\right)$
and $\left(t'_2,{\bf x}'_2\right) = \left(t_2,{\bf x}_2\right)$
from the start. This will be confirmed during the computation
of the vacuum expectation value.

First we consider just the square of the energy density operator; its
expectation value is
\beqn
\left<0\left|\hat\rho^2\right|0\right> 
%
%
&=& {\frac{1}{4}} \int d\mu\left({\bf k}_1,{\bf k}_2\right) 
{{N^2_{k_{1}}}}\,{{N^2_{k_{2}}}}\;
{e^{-2\,{{\sigma}^2}\,\left( {{k^2_{1}}} + {{k^2_{2}}} \right) }}
\left\{
\right.\nonumber \\
&&\hspace{5mm}
{{\left( {\bf k}_{1}\cdot{\bf k}_{2} + \omega_{1}\,{\omega_2} \right) }^2}
  \left(
{e^{i\,\left( {\bf k}_{1}\cdot\left({\bf x}_{1} - {{\bf x}'_2}\right) + 
        {\bf k}_{2}\cdot\left({{\bf x}'_1} - {{\bf x}_2}\right) \right)  - 
     i\,\left( \left(t_{1} - {t'_2}\right)\,\omega_{1} + 
        \left( {t'_1} - {t_2} \right) \,\omega_{2} \right) }}
\right.\nonumber \\
&&\hspace{31mm}\left.
 + {e^{i\,\left( {\bf k}_{1}\cdot\left({\bf x}_{1} - {{\bf x}_2}\right) + 
        {\bf k}_{2}\cdot\left({{\bf x}'_1} - {{\bf x}'_2}\right) \right)  - 
     i\,\left( \left(t_{1} - {t_2}\right)\,\omega_{1} + 
        \left( {t'_1} - {t'_2} \right) \,\omega_{2} \right) }}
  \right)
\nonumber \\
&&\hspace{2mm}
+ \left[
   {{\omega^2_{1}}}
   \left( {e^{-i\,\left( {\bf k}_{1}\cdot\left({\bf x}_{1} - {\bf 
x}_{2}\right) - 
        \left( t_{1} - {t'_1} \right) \,\omega_{1} \right) }} + 
  {e^{i\,\left( {\bf k}_{1}\cdot\left({\bf x}_{1} - {{\bf x}'_1}\right) -
        \left( t_{1} - {t'_1} \right) \,\omega_{1} \right) }} \right)
 \right.
\nonumber \\
&&\hspace{3mm}\left.\times
\left.
{{\omega^2_{2}}}
   \left( {e^{-i\,\left( {\bf k}_{2}\cdot\left({{\bf x}_2} - {\bf 
x}_{4}\right) -
        \left( {t_2} - {t'_2} \right) \,\omega_{2} \right) }} + 
  {e^{i\,\left( {\bf k}_{2}\cdot\left({{\bf x}_2} - {{\bf x}'_2}\right) -
        \left( {t_2} - {t'_2} \right) \,\omega_{2} \right) }} \right)
 \right]
\right\}
\eeqn
By comparing the last two lines of the above expression with
Eq.(\ref{energy-density-general}), we see this is but
$\rho\left(t_1,{\bf x}_1; t'_1,{\bf x}'_1;\sigma\right)
\rho\left(t_2,{\bf x}_2; t'_2,{\bf x}'_2;\sigma\right)$. Thus, the
remainder is the desired expression for
$\Delta\rho^2(t_1,{\bf x}_1,t'_1,{\bf x}'_1;t_2,{\bf x}_2,t'_2,{\bf x}'_2;\sigma)$.
Even for the case $\sigma \rightarrow 0$, this expression is
finite for $(t'_1,{\bf x}'_1)\rightarrow (t_1,{\bf x}_1)$ and
$(t'_2,{\bf x}'_2)\rightarrow (t_2,{\bf x}_2)$, as long as
$(t_1,{\bf x}_1) \ne (t_2,{\bf x}_2)$. Letting
$(t,{\bf x}) = (t_2,{\bf x}_2) - (t_1,{\bf x}_1)$, our results
for the energy density [from (\ref{energy-density-general})] and its correlation function 
[from here] are

\begin{equation}
\rho\left(t,{\bf x};\sigma\right) = \int d\mu\left({\bf k}\right) 
  {{N^2_{k}}}\,{{\omega^2}}\, {e^{-2\,{{\sigma}^2}\,{{k^2}}}}\, 
  \cos ({\bf x}\cdot{\bf k} -  t\,\omega)
\label{rho-xsigma}
\end{equation}

\begin{equation}
\Delta\rho^2\left(t,{\bf x};\sigma\right) =
{\frac{1}{2}}
\int d\mu\left({\bf k}_1,{\bf k}_2\right) 
{{N^2_{k_{1}}}}\,{{N^2_{k_{2}}}}\,
  {{\left( {\bf k}_{1}\cdot{\bf k}_{2} + \omega_{1}\,\omega_{2} \right) }^2}\,
{e^{
 - 2\,{{\sigma}^2}\,\left( {{k^2_{1}}} + {{k^2_{2}}} \right) 
 - i\,{\bf x}\cdot\left({\bf k}_{1} + {\bf k}_{2}\right)
 + i\,t\,\left( \omega_{1} + \omega_{2} \right) }}
\label{Drho-xsigma}
\end{equation}

\section{Smeared-Field Energy Density and Fluctuations in Minkowski Space}

We consider a  Minkowski space $R^1 \times R^d$ with $d$-spatial dimensions. For this 
space the mode density is
\beq
\int d\mu({\bf k})  = \int_0^\infty k^{d-1}\,dk\int_{S^{d-1}} d\Omega_{d-1}
\quad{\rm with}\quad
\int_{S^{d-1}} d\Omega_{d-1} = {\frac{2\,{{\pi 
}^{{\frac{d}{2}}}}}{\Gamma\!\left({\frac{d}{2}}\right)}}
\eeq
and the mode function normalization constant is
$N_{k_{1}} = 1/{\sqrt{{2^{d+1}}\,{{\pi }^d}\,\omega_{1}}}$.
We introduce the angle between two momenta in phase space, $\gamma$, via
\beq
{\bf k}_1 \cdot {\bf k}_2 = k_1 k_2 \cos(\gamma)
                          = \omega_1 \omega_2 \cos(\gamma).
\eeq
The averages of the cosine and cosine squared of this angle over a pair of unit
spheres are
\begin{mathletters}
\beqn
\int_{S^{d-1}} d\Omega_1 \int_{S^{d-1}} d\Omega_2 \cos(\gamma) &=& 0 \\
\int_{S^{d-1}} d\Omega_1 \int_{S^{d-1}} d\Omega_2 \cos^2(\gamma) &=& 
       {\frac{4\,{{\pi }^d}}{d\,{{\Gamma\!\left({\frac{d}{2}}\right)}^2}}}.
\eeqn
\end{mathletters}
The smeared energy density (\ref{rho-sigma}) becomes 
\beqn
\rho(\sigma) &=& 
{\frac{1}{{2^d}\,{{\pi }^{{\frac{d}{2}}}}\,\Gamma\!\left({\frac{d}{2}}\right)}}
\int_0^\infty {\frac{{{k_{1}}^d}}{{e^{2\,{{\sigma}^2}\,{{k_{1}}^2}}}}} \,dk_1
\nonumber \\
&=&
\frac{
  \Gamma\!\left({\frac{d+1}{2}}\right)
     }{
  {2^{{\frac{3\,\left( d+1 \right) }{2}}}}\,{{\pi }^{{\frac{d}{2}}}}\,
     \sigma^{d+1} \Gamma\!\left({\frac{d}{2}}\right)
     }
\eeqn

For the fluctuations of the smeared energy density operator, 
we evaluate (\ref{Drho-xsigma})  for this space and find
\beqn
\Delta\rho^2(\sigma) &=&
{\frac{1}{{2^{(2d+3)}}{{\pi }^{2\,d}}}}
\int_0^\infty \int_0^\infty
\int_{S^{d-1}}\int_{S^{d-1}}
{\frac{{{\left( 1 + {\cos(\gamma)} \right) }^2}\,{{k^d_{1}}}\,{{k^d_{2}}}}
   {{e^{2\,{{\sigma}^2}\,\left( {{k^2_{1}}} + {{k^2_{2}}} \right) }}}} 
\,d\Omega_1 \,d\Omega_2 \,dk_1\,dk_2
\nonumber \\
&=&
{\frac{\left( d+1 \right) \,
     {{\Gamma\!\left({\frac{d+1}{2}}\right)}^2}}{{2^{(3d+4)}}\,d\,{{\pi }^d}\,
     {{\sigma}^{2\,\left( d+1 \right) }}\,
     {{\Gamma\!\left({\frac{d}{2}}\right)}^2}}}
\eeqn
Defining the (dimension-dependent) constant
\beq
\chi_d \equiv 
{\frac{1+d}{2\,d}},
\eeq
we write the smeared fluctuations in terms of the square of the
smeared energy density
\beq
\Delta\rho^2(\sigma) = \chi_d\; \rho(\sigma)^2
\eeq
We introduce the dimensionless measure of fluctuations
\beq
\Delta = \left|
  1 - \frac{\left<\rho\right>^2}{\left<\rho^2\right>}
           \right|
= \left|\frac{\Delta\rho^2}{\Delta\rho^2 + \left<\rho\right>^2} \right|
= \frac{\chi_d}{\chi_d+1}
\eeq
and for Minkowski space we have
\beq
\Delta_{\rm Minkowski}(d) = 
  {\frac{1+d}{1 + 3\,d}} 
\eeq
which has the particular values
\[
\begin{array}{|c||c|c|c|c|}
\hline
d & 1 & 3 & 5 & \infty \\ \hline
\Delta_{\rm Minkowski} & \frac{1}{2} & \frac{2}{5} & \frac{3}{8} & \frac{1}{3} 
\\ \hline
\end{array}
\]

\section{Smeared-Field in Casimir Topology}
The Casimir topology is obtained from a  flat space (with $d$ spatial dimensions, i.e.,  
$R^1 \times R^{d}$ ) by imposing  periodicity $L$ in one of its spatial dimensions, 
say,  $z$, thus endowing it  with
a  $R^1\!\times\!R^{d-1}\!\times\!S^1$  topology. 
We decompose ${\bf k}$ into a component along the  periodic dimension and 
call the remaining components ${\bf k}_{\perp}$:
\begin{mathletters}
\label{Cdu}
\begin{eqnarray}
{\bf k} &=& \left({\bf k}_\perp,\frac{2 \pi n}{L}\right) 
        = \left({\bf k}_\perp, l n\right), l \equiv 2 \pi/L \\ 
\omega_{1} &=& {\sqrt{{{k^2_{1}}} + {l^2}\,{{n_{1}}^2}}} 
\end{eqnarray}
The normalization and momentum measure are
\begin{eqnarray}
\int d\mu({\bf k}) &=&\int_0^\infty k^{d-2}\,dk\int_{S^{d-2}} d\Omega_{d-2}
\sum_{n=-\infty}^\infty \\
N_{k_{1}} &=& {\frac{1}{{\sqrt{{2^d}\,L\,{{\pi }^{d-1}}\,\omega_{1}}}}}
\end{eqnarray}
\end{mathletters}
With this, the energy density (\ref{rho-sigma}) becomes
\beqn
\rho_L\left(\sigma\right) &=&
{\frac{l}
   {{2^d}\,{{\pi }^{{\frac{d+1}{2}}}}\,\Gamma\!\left({\frac{d-1}{2}}\right)}}
\sum_{n_1=-\infty}^\infty\int_0^\infty 
{{k_{1}^{d-2}}}\,\left({{k^2_{1}}} + {l^2}\,{n^2_1}\right)^{\frac{1}{2}}\,
{e^{-2\,{{\sigma}^2}\,\left( {{k^2_{1}}} + {l^2}\,{n^2_1} \right) }}\;dk_1
\nonumber \\
&=&
{\frac{l}
   {{2^d}\,{{\pi }^{{\frac{d+1}{2}}}}\,\Gamma\!\left({\frac{d-1}{2}}\right)}}
\sum_{n_1=-\infty}^\infty\int_0^\infty 
{\frac{{{k_{1}}^d}}{{\sqrt{{{k^2_{1}}} + {l^2}\,{n^2_1}}}}} 
{e^{-2\,{{\sigma}^2}\,\left( {{k^2_{1}}} + {l^2}\,{n^2_1} \right) }} \;dk_1
\nonumber \\
&&+
{\frac{l}
   {{2^d}\,{{\pi }^{{\frac{d+1}{2}}}}\,\Gamma\!\left({\frac{d-1}{2}}\right)}}
\sum_{n_1=-\infty}^\infty\int_0^\infty 
{\frac{{l^2}\,{{k_{1}}^{-2 + d}}\,{n^2_1}}
   {{\sqrt{{{k^2_{1}}} + {l^2}\,{n^2_1}}}}} {e^{-2\,{{\sigma}^2}\,\left( 
{{k^2_{1}}} + {l^2}\,{n^2_1} \right) }} \;dk_1
\eeqn
Using the analysis of the Appendix A we write this as
the sum of the two smeared Green function derivatives
\beqn
\rho_L\left(\sigma\right) &=& 
      \left<0_L\left| \left(
        \left( \nabla_{\!\perp} \phi_t \right)\left(f_{\bf x}\right)
      \right)^2 \right|0_L\right>
+
      \left<0_L\left| \left(
        \left( \partial_z \phi_t \right)\left(f_{\bf x}\right)
      \right)^2 \right|0_L\right>
\nonumber \\
&=& \Gperp + \Gz
\eeqn
where $\left.\left.\right|0_L\right>$ is the Casimir vacuum.

\subsection{Regularized Casimir Energy Density}

Since ${G_L(\sigma)}_{,i} = G_{L,i}^{\rm div} + G_{L,i}^{\rm fin}$
($i=x_\perp x_\perp$ or $zz$)
we see how to split the smeared energy density into a
$\sigma\rightarrow 0$ divergent term and the finite contribution:
\beq
\rho_L\left(\sigma\right) = \rho_L^{\rm div} + \rho_L^{\rm fin}
\eeq
where
\beqn
\rho_L^{\rm div} &=& \Gperpdiv +\Gzdiv
\nonumber \\ &=&
{\frac{\,\Gamma\!\left({\frac{d+1}{2}}\right)}
   {{2^{{\frac{3\,\left( d+1 \right) }{2}}}}\,{{\pi }^{{\frac{d}{2}}}}\,
     {{\sigma}^{d+1}}\Gamma\!\left({\frac{d}{2}}\right)}}
\nonumber \\
&=& \rho\left(\sigma\right)
\eeqn
and
\beqn
\rho_L^{\rm fin} &=& \Gperpfin + \Gzfin
\nonumber \\
&=&  
-\frac{
   d\,
\Gamma\!\left(-{\frac{d}{2}}\right)\Gamma\!\left({\frac{d}{2}}\right)\,
      }{
   (4\pi)^{(d+3)/2}\,{l^{d+1}}\, 
}
\sum_{p=1}^\infty
{{\left( -1 \right) }^p}\,
 {(2\,l)^{2p}}\,
  p\, {{\left( 2p-1 \right) }^2}\,
  {{\sigma}^{2\,\left(p-1\right) }} 
{\frac{B_{2p+d-1}}{2p+d-1}} 
{\frac{\left( 2p-3 \right) !!}{\left( 2\,p \right) !}} 
{\frac{
     \Gamma\!\left(p-{\frac{1}{2}}\right)}{\Gamma\!\left(p+{\frac{d}{2}}
     \right)}}
\eeqn
With this we define the regularized energy density
\beqn
\rho_{L,{\rm reg}} &\equiv& \lim_{\sigma\rightarrow 0}\left(
 \rho_L\left(\sigma\right) - \rho\left(\sigma\right)
\right)\nonumber \\
&=&  {\frac{d\,\,{{\pi }^{{\frac{d}{2}}}}\,B_{d+1}\,
     \Gamma\!\left(-{\frac{d}{2}}\right)\,\Gamma\!\left({\frac{d}{2}}\right)}
     {2\,\left( d+1 \right) {L^{d+1}}\,\Gamma\!\left({\frac{d}{2}}+1\right)}}
\eeqn
and get the usual results
\[
\begin{array}{|c||c|c|c|}
\hline
     d             &          1 & 3 & 5 \\ \hline 
\rho_{L,{\rm reg}} &  - {\frac{{{\pi }}}{6\,{L^2}}} &
 - {\frac{{{\pi }^2}}{90\,{L^4}}} &
          - {\frac{2\,{{\pi }^3}}{945\,{L^6}}} 
\\ \hline
\end{array}
\]

\subsection{Casimir energy density fluctuations}

For the $d$-dimensional Casimir geometry, the fluctuations are
\beqn
\Delta\rho^2_L\left(\sigma\right) &=&
{\frac{{l^2}}{{2^{2d+3}}\,{{\pi }^{2\,d}}}}
\sum_{n_1=-\infty}^\infty \sum_{n_2=-\infty}^\infty
\int_0^\infty \hspace{-3mm}k_1^{d-2}\,dk_1 \int_0^\infty \hspace{-3mm}k_2^{d-2}\,dk_2
\int_{S^{d-2}}\hspace{-5mm} d\Omega_{1} \int_{S^{d-2}}\hspace{-5mm} d\Omega_{2}
\nonumber \\ && \hspace{10mm}\times
\frac{
{e^{-2\,{{\sigma}^2}\left(\omega_1^2 + \omega_2^2\right)}} 
}{\omega_{1}\,\omega_{2}}
\left( {\cos(\gamma)}\,k_{1}\,k_{2} + {l^2}\,n_{1}\,n_{2} + 
         \omega_{1}\,\omega_{2} \right)^2
\nonumber \\
&=& 
{\frac{{l^2}}
   {{2^{2\,d}}\,{{\pi }^{d+1}}\,{{\Gamma\!\left({\frac{d-1}{2}}\right)}^2}}}
\sum_{n_1,n_2=-\infty}^\infty
\int_0^\infty \!\! \int_0^\infty \!\! dk_1 \,dk_2 \,
 \frac{k_1^{d+2}e^{-2\,\sigma^2\,\left(k_1^2 + l^2\,n_1^2 \right)}}
         {\sqrt{{k_1^2} + {l^2}\,{n_1^2}}}
 \frac{k_2^{d+2}e^{-2\,\sigma^2\,\left(k_2^2 + l^2\,n_2^2 \right)}}
         {\sqrt{{k_2^2} + {l^2}\,{n_2^2}}}
\nonumber \\&&\hspace{10mm}\times
\left(
  {\frac{d\,{k_1^2}\,{k_2^2}}{2\,\left( d-1 \right) }} 
+ {\frac{l^2}{2}}\left( {k_2^2}\,{n_1^2} + 
{k_1^2}\,{n_2^2} \right) 
+ {l^4}\,{n_1^2}\,{n_2^2}
\right)
\eeqn
We write this expression in terms of products of
the Green functions derivatives used above:
\beq                 
\Delta\rho^2_L\left(\sigma\right) = 
  {\frac{d\,\left(\Gperp\right)^2
      }{2\,\left( d-1 \right) }} 
+  
    \Gz
       \left(\Gperp  + \Gz \right)
\eeq
We can
split $\Delta\rho^2_L\left(\sigma\right)$ into three general terms
\beq
\Delta\rho^2_L\left(\sigma\right) = \Delta\rho^{2,{\rm div}}_L
   + \Delta\rho^{2,{\rm cross}}_L
   + \Delta\rho^{2,{\rm fin}}_L
\eeq
The first term contains only the divergent parts of the Green functions
while the last term contains only the finite parts. This is similar
to the split we used for the smeared energy density above. What is new
here is the middle term $\Delta\rho^{2,{\rm cross}}_L$. This comes about
from the products of the divergent part of one Green function and the
finite part of the other. That this term arises for computations of the
energy density fluctuations is a generic feature. We will discuss in greater detail
the meaning of this term later.

The results of Appendix A give
\begin{mathletters}
\beqn
\Delta\rho^{2,{\rm div}}_L &=& 
 {\frac{d\,\left(\Gperpdiv\right)^2 }
       {2\,\left( d-1 \right) }} 
      + \Gzdiv\left(\Gperpdiv+H_2^{\rm div}\right)
\nonumber \\
&=&
{\frac{\left( d+1 \right) \,
     {{\Gamma\!\left({\frac{d+1}{2}}\right)}^2}}{d\,{2^{3\,d+4}}\,{{\pi }^d}\,
     {{\sigma}^{2\,\left( d+1 \right) }}\,
     {{\Gamma\!\left({\frac{d}{2}}\right)}^2}}}
\nonumber \\
&=& \chi_d\left( \rho_L^{\rm div} \right)^2
=  \chi_d\left( \rho\left(\sigma\right)\right)^2
\eeqn
\beqn
\Delta\rho^{2,{\rm cross}}_L &=& 
 {\frac{d}{d-1}} \Gperpdiv \Gperpfin
      + 
  \left(
    2\Gzdiv \Gzfin
    + \Gperpdiv \Gzfin
    +\Gzdiv \Gperpfin
  \right)
\nonumber \\
&=&
{\frac{\left( d+1 \right) \,
     \Gamma\!\left({\frac{d+1}{2}}\right)}{{2^
       {{\frac{3\,\left( d+1 \right) }{2}}}}\,d\,{{\pi }^{{\frac{d}{2}}}}\,
       {{\sigma}^{d+1}}\,
     \Gamma\!\left({\frac{d}{2}}\right)}} \left( \Gperpfin + \Gzfin 
\right)
\nonumber \\
&=&
2 \chi_d \, \rho_L^{\rm div} \, \rho_L^{\rm fin}
\eeqn
\beqn
\Delta\rho^{2,{\rm fin}}_L &=& 
 {\frac{d\, \left(\Gperpfin\right)^2 }
       {2\,\left( d-1 \right) }}
      +  \Gzfin\left(\Gperpfin+\Gzfin\right)
\nonumber \\
&=& 
\frac{d^2\,l^{2(d+1)}}{2^{2d+7}\,\pi^{d+3}}
  {{\Gamma\!\left(-{\frac{d}{2}}\right)}^2}\,
  {{\Gamma\!\left( {\frac{d}{2}}\right)}^2}
\nonumber \\&&\times
\sum_{p,q=1}^\infty \left(
{{\left( -1 \right) }^{p + q}}\,{2^{2\,\left( p + q \right) }}\,
  {l^{2\,\left( p + q \right) }}\,p\,q\,\left( 2\,p -1\right) \,
  \left( 2\,q -1 \right) \,{{\sigma}^{2\,\left( p + q-2 \right) }}
\right.\nonumber \\&&\hspace{10mm}\times\left.
\left( 4 + {d^2} - 6\,q + d\,\left( 2\,p + 2\,q - 3 \right)  + 
  2\,p\,\left( 4\,q - 3\right)  \right)
\right.\nonumber \\&&\hspace{10mm}\times\left.
{\frac{B_{2\,p+d-1}\,B_{2\,q+d-1}\,\left( 2\,p - 3\right) !!\,
     \left( 2\,q - 3\right) !!}{\left( 2\,p + d - 1\right) \,
     \left( 2\,q + d - 1\right) \,\left( 2\,p \right) !\,
     \left( 2\,q \right) !}} {\frac{\Gamma\!\left(p-{\frac{1}{2}}\right)\,
     \Gamma\!\left(q-{\frac{1}{2}}\right)}{\Gamma\!\left(p+{\frac{d}{2}}
      \right)\,\Gamma\!\left({q + \frac{d}{2}}\right)}}
\right)
\nonumber \\
&\stackrel{\sigma\rightarrow 0}{\longrightarrow}&
{\frac{{d^3}\,{{\pi }^d}\,{B_{d+1}^2}\,
     {{\Gamma\!\left(-{\frac{d}{2}}\right)}^2}\,
     {{\Gamma\!\left({\frac{d}{2}}\right)}^2}}{8\,\left( d+1 \right) \,
     {L^{2\,\left( d+1 \right) }}\,
     {{\Gamma\!\left(1 + {\frac{d}{2}}\right)}^2}}}
\nonumber \\
&=&
{\frac{d\,\left( d+1 \right) }{2}}
 \left( \rho_{L,{\rm reg}} \right)^2
\eeqn 
\end{mathletters}

From this we see the divergent and cross terms can be related to
the smeared energy density via
\beq
\Delta\rho^{2,{\rm div}}_L + \Delta\rho^{2,{\rm cross}}_L = 
 \chi_d\left\{
    \left(\rho_L^{\rm div}\right)^2 + 2 \rho_L^{\rm div} \rho_L^{\rm fin}
\right\}
\eeq
where $\chi_d$ is the function that relates the fluctuations of the
energy density to the mean energy density when the boundaries are not
present, i.e., Minkowski space. This leads us to interpret these terms 
as due the vacuum fluctuations that are always present. With this in mind, 
we define the regularized fluctuations of the energy density
\beqn
\Delta\rho^2_{L,{\rm reg}} &=&
 \lim_{\sigma\rightarrow 0}\left(
      \Delta\rho^2_L\left(f\right)
    - \chi_d\left\{
      \left(\rho_L^{\rm div}\right)^2 + 2 \rho_L^{\rm div} \rho_L^{\rm fin}
      \right\}
 \right)
\nonumber \\
&=& \chi_{d,L} \left( \rho_{L,{\rm reg}} \right)^2
\eeqn
where 
\beq
\chi_{d,L} \equiv {\frac{d\,\left( d+1 \right) }{2}}.
\eeq
We also define a regularized version of the dimensionless measure
$\Delta$:
\beq
\Delta_{L,{\rm Reg}} \equiv
  \frac{\Delta\rho^2_{L,{\rm Reg}}}
       {\Delta\rho^2_{L,{\rm Reg}} + \left(\rho_{L,{\rm Reg}}\right)^2}
=
{\frac{d\,\left( d+1 \right) }{2 + d + {d^2}}} 
\eeq
and note the values:
\[
\begin{array}{|c||c|c|c|c|}
\hline
d & 1 & 3 & 5 & \infty \\ \hline 
\Delta_{L,{\rm Reg}} & \frac{1}{2} & \frac{6}{7} & \frac{15}{16} & 1
\\ \hline
\end{array}
\]

Following the procedures described in Appendix B, we have made two plots,
Fig. 1 of $\Delta (\sigma, L)$ and $\Delta_{L, Reg}$ versus $\sigma /L$, (which 
we call  $\sigma'$ here for short); 
and Fig. 2 of $\rho_{L, Reg}$ and $\sqrt {\Delta \rho^2_{L, Reg}}$ versus $\sigma'$.
The range of $\sigma'$ is limited to $\le 0.4$ because going 
any further would make the meaning of a local energy density ill-defined, as  
the smearing of the field extends to the Casimir boundary in  space.
(We believe  this infrared limit also carry important physical meaning
in reference to the structure of spacetime,  it is outside the focus of this paper.)

Let us ponder on the meaning they convey.  In Fig. 1, we first note that both curves 
are of the order unity. But the behavior of $\Delta$ (recall that the energy density 
fluctuations thus  defined include the cross term along with 
the finite part and the  state independent divergent part)  
is relatively insensitive to the smearing width, whereas $\Delta_{L,Reg}$,
which measures only the finite part of the energy density fluctuations to
the mean has more structure.   In particular, it saturates its upper bound of 1
around $\sigma' = 0.24$. Note that if one adheres to the KF criterion \cite{KuoFor}
one would say that semiclassical gravity fails, but all that is happening here
is that $\rho_{L,{\rm Reg}}=0$ while $\Delta \rho_{L,{\rm Reg}}^2 $ shows no special 
feature.  The real difference between 
these two functions is the cross term, which is responsible for their 
markedly different structure and behavior.
We have more to say about what to make of the cross term in the 
last section, which should be contrasted with the opinion
of  Wu and Ford \cite{WuFor} on its physical significance.
In Fig. 2, the main feature to notice is that the regularized  energy density 
crosses from negative to positive values at around $\sigma' = 0.24$ . The negative
Casimir energy density  calculated in a point-wise field theory which corresponds
to small ranges of $\sigma'$ is expected, and is usually taken to
signify the quantum nature of the Casimir state. As  $\sigma'$ increases
we are averaging the field operator over a larger region, and thus sampling
the field theory from the ultraviolet all the way to the infrared region. At large
$\sigma'$ finite size effect begins to set in. The difference and relation 
of these two effects are explained in  \cite{HuO'C}: Casimir effect
arises from summing up the quantum fluctuations of ALL modes
(as altered by the boundary), with no insignificant short wavelength 
contributions, whereas finite size effect has dominant contributions from 
the LONGEST wavelength modes, and thus reflect the large scale 
behavior.  As the smearing moves from a small scale to the far boundary of space,
the behavior of the system is expected to shift from a Casimir-dominated to 
a finite size-dominated effect.  This could be the underlying reason in 
the  crossover behavior of $\rho_{L, Reg}$. 


\section {Point Separated Energy Density and Fluctuations in Minkowski Space}

We return now to the Minkowski space and consider the point-separated expressions
for the energy density and its fluctuations. This will lead to some interesting new
observations about point-separation and maybe even the extended structure of spacetime.
Consider the point separated energy density (\ref{rho-xsigma}) with $\sigma = 0$ in 
a Minkowski space with $d$-spatial dimensions
\beqn
\rho(t,{\bf x}) &=& 
\int d\mu({\bf k}) \,N_k^2 \, \omega^2 \,
\cos ({\bf x}\cdot{\bf k} - t\,\omega)
\nonumber \\
&=& {\frac{1}
   {{2^d}\,{{\pi }^{{\frac{d+1}{2}}}}\Gamma\!\left({\frac{d-1}{2}}\right)}}
\int_0^\infty \int_{-\infty}^\infty
\cos\left(x\,k_{x} - t\,{\sqrt{ k_\perp^2 + k_x^2}}\right)\,
  k_\perp^{d-2}\,
  \sqrt{k_\perp^2 + k_x^2} \, dk_x\,dk_\perp
\eeqn
where we take ${\bf x} = {\bf x}_1 - {\bf x}_2 = x \hat {x}$
and decompose $ {\bf k} = ( k_{x}, {\bf k}_{\perp})$ into one
component along $\hat x$ and two perpendicular to $\hat x$.
We change  variables to  $k_x = k\cos\phi$ and $k_\perp = |{\bf k}_\perp| = k\sin\phi$
and evaluate
\beqn
\rho(t,x) &=& {\frac{1}
   {{2^d}\,{{\pi }^{{\frac{d+1}{2}}}}\,\Gamma\!\left({\frac{d-1}{2}}\right)}}
\int_0^\infty \int_0^\pi
{k^d}\,\cos (k\,\left( t - x\,\cos\phi \right) )\,
  \sin^{d+2}\!\phi
\;d\phi\,dk
\nonumber \\
&=& \frac{1}{2^{\frac{d}{2}+1}\, \pi^\frac{d}{2}}
\int_0^\infty
{k^d}\,{{\left( {k^2}\,{x^2} \right) }^{{\frac{2 - d}{4}}}}\,
  J_{{\frac{d}{2}}-1}\left({\sqrt{{k^2}\,{x^2}}}\right)\,\cos (k\,t) \;dk
\nonumber \\
&=& \frac{1}{2^{\frac{d}{2}+1}\, \pi^\frac{d}{2}}
\Re\int_0^\infty {k^d}\,
  {{\left( {k^2}\,{x^2} \right) }^{\frac{2-d}{4}}\,
  J_{{\frac{d}{2}}-1}}\left({\sqrt{{k^2}\,{x^2}}}\right)
  {e^{-\epsilon\,k  + i\,k\,t}}
\;dk
\eeqn
A small positive imaginary part $i\epsilon$ is added to $t$ to
guarantee the $k$ integral convergences. Continuing,
\beqn
&=&
{\frac{1}{2\pi^{\frac{d+1}{2}}}} \Re\left\{
\frac{
  \left( d\,{{\left( \epsilon - i\,t \right) }^2} - {x^2} \right)
     }{
   {{\left( \epsilon - i\,t \right) }^{d+3}}
     } \,
  {{\left( 1 + {\frac{{x^2}}{{{\left( \epsilon - i\,t \right) }^2}}} \right)
       }^{-{\frac{d+3}{2}}}}\,\Gamma\!\left({\frac{d+1}{2}}\right)
\right\}
\nonumber \\
&=&
-{\frac{1}{2\pi^{\frac{d+1}{2}}}}
\frac{
    \left( d\,{t^2} + {x^2} \right)
     }{
   {{\left( {t^2} - {x^2} \right) }^{{\frac{d+3}{2}}}}
     }
  \Gamma\!\left({\frac{d+1}{2}}\right)\,\sin ({\frac{d\,\pi }{2}})
\eeqn
The $\sin\left(d\pi/2\right)$ factor shows this result only holds for
odd $d$. Restricting ourselves to odd $d$'s, the final result for the
point separated energy density in Minkowski space is
\beq
\rho(t,x) =
-{\frac{ \left( -1 \right)^\frac{d-1}{2}  
    \Gamma\!\left({\frac{d+1}{2}}\right)
}{2\pi^{\frac{d+1}{2}}}}
\frac{
    \left( d\,{t^2} + {x^2} \right)
     }{
   {{\left( {t^2} - {x^2} \right) }^{{\frac{d+3}{2}}}}
     }
\eeq

Now we consider the correlation function (\ref{Drho-xsigma}).
 By identifying  the two point function derivatives
\begin{mathletters}
\beqn
G_{x_\perp x_\perp}(t,x) &=& 
-\left<0\left|
   \nabla_{x_{\!\perp}}^2 \!\!\left(\hat\phi(t_1,{\bf x}_1)\hat\phi(t_2,{\bf x}_2)\right)
\right|0\right>
\nonumber \\
&=&
\frac{1}{2^d\,\pi^\frac{d+1}{2}\,\Gamma\!\left(\frac{d-1}{2}\right)}
\int_0^\infty \int_{-\infty}^\infty
   {\frac{{k_\perp^d}\,{e^{-i\,\left( x\,k_x - 
           t\,{\sqrt{{k_\perp^2} + { k_x ^2}}} \right) }}}
   {{\sqrt{{k_\perp^2} + {k_x^2}}}}} \;dk_x\,dk_\perp
\\ \nonumber \\
G_{xx}(t,x) &=& 
-\left<0\left|
   \frac{\partial^2}{\partial x^2}\!\left(\hat\phi(t_1,{\bf x}_1)\hat\phi(t_2,{\bf x}_2)\right)
\right|0\right>
\nonumber \\
&=&
\frac{1}{2^d\,\pi^\frac{d+1}{2}\,\Gamma\!\left(\frac{d-1}{2}\right)}
\int_0^\infty \int_{-\infty}^\infty
   {\frac{{k_\perp^{d-2}}\,{k_x^2} }
   {
     {\sqrt{{k_\perp^2} + {k_x^2}}}}}
{e^{-i\,\left( x\,k_x - 
           t\,{\sqrt{{k_\perp^2} + {k_x^2}}} \right) }}
\; dk_x \, dk_\perp
\\ \nonumber \\
G_{tx}(t,x) &=& 
\left<0\left|
   \frac{\partial^2}{\partial t\,\partial x} \!
   \left(\hat\phi(t_1,{\bf x}_1)\hat\phi(t_2,{\bf x}_2)\right)
\right|0\right>
\nonumber \\
&=&
\frac{1}{2^d\,\pi^\frac{d+1}{2}\,\Gamma\!\left(\frac{d-1}{2}\right)}
\int_0^\infty \int_{-\infty}^\infty
   {k_\perp^{d-2}}\,k_x \,
   e^{-i\,\left( x\,k_x - 
          t\,{\sqrt{{k_\perp^2} + {k_x^2}}} \right) }
\; dk_x \, dk_\perp
\eeqn
\end{mathletters}
we write the energy density correlation function as
\beq
\Delta\rho^2(t,x) =
{\frac{d\,{G_{x_\perp x_\perp}^2(t,x)}}{2\,\left( d-1 \right) }} + 
{G_{x_\perp x_\perp}(t,x)}\,G_{xx}(t,x) +
  G_{xx}^2(t,x) + G_{tx}^2(t,x)
\eeq
We proceed with the evaluation of the Green functions in a similar manner
as we did for the point separated energy density above and obtain
%
%
\begin{mathletters}
\beqn
G_{x_\perp x_\perp}(t,x) &=& 
\frac{
    (-1)^\frac{d+1}{2}
    \Gamma\!\left({\frac{d+1}{2}}\right)
      }{
    2\, \pi^\frac{d+1}{2}
      }
\frac{
        (d-1)
     }{
         \left( t^2 - x^2 \right)^{\frac{d+1}{2}}     
     }
\\
%
%
G_{xx}(t,x) &=& 
\frac{
   (-1)^\frac{d+1}{2} \,\Gamma\!\left({\frac{d+1}{2}}\right)
      }{
   2 \, \pi^\frac{d+1}{2}
      }
\frac{
  \left( {t^2} + d\,{x^2} \right)
     }{
  \left( {t^2} - {x^2} \right)^{\frac{d+3}{2}}
     }
\\
%
%
G_{tx}(t,x) &=& 
\frac{
   (-1)^\frac{d+1}{2} \, 
        \Gamma\!\left({\frac{d+1}{2}}\right)}{{2\, \pi^\frac{d+1}{2}}}
\frac{
   (d+1)  \, t\,x
     }{
  {{\left({x^2} - {t^2} \right) }^{{\frac{d+3}{2}}}}\,
     }
\eeqn
\end{mathletters}
With these results, correlation function is
\beq
\Delta\rho^2(t,x) =
\frac{{{\Gamma\!\left({\frac{d+1}{2}}\right)}^2}}{ \pi^{d+1}}
\left(\frac{
4\,{t^2}\,{x^2} + d\,{{\left( {t^2} + {x^2} \right) }^2} 
     }{
   {{\left( {t^2} - {x^2} \right) }^{d+3}}\,
     }\right)
\eeq
The constant $\chi_d$ is now a function of the temporal and
spatial separation,
\beq
\chi_d(t,x) =
\frac{d+1}{2}\left(
\frac{
   4\,{t^2}\,{x^2} +  d\,{{\left( {t^2} + {x^2} \right) }^2}
     }{
     \left( d\,\,{t^2} + {x^2} \right)^2
     }
\right),
\eeq
and we write the correlation function in terms of the square of
the point separated energy density
\beq
\Delta\rho^2(t,x) = \chi_d(t,x) \left( \rho(t,x) \right)^2
\eeq
Our dimensionless measure is  also a function of the
separation:
\beq
\Delta(t,x) =
\frac{
    \left( d+1 \right) \,\left(
	d\,{(t^2+x^2)}^2 + 4\,t^2\,x^2
    \right) 
     }{
    \left( d+1 \right) \,\left(
	d\,{(t^2+x^2)}^2 + 4\,t^2\,x^2
    \right) 
    + 2\,{(d\,t^2+x^2)}^2
     }
\eeq
To extract physical  meaning out of this for a point-wise quantum field theory, 
we have to work in the $(t,x)\rightarrow 0$ limit 
(recall $t = t_1 - t_2$, ${\bf x} = {\bf x}_1 - {\bf x_2} = x\hat x$), 
for only then 
$\rho(t,x)\rightarrow \left<0\left|\hat\rho\right|0\right>$. With
this in mind, we parameterize the direction dependence via
$t=r\sin\theta$ and $x = r\cos\theta$ (note this is only a parameterization, the
imaginary time $\tau$ and $x$ shown below would carry physical meaning 
in the Euclidean sense). In the  $r\rightarrow 0$ limit we have
\beq
\Delta(\theta) = 
\frac{
  \left( d+1 \right) \,\left( 1 + 2\,d - \cos (4\,\theta) \right) 
     }{
  \left( d+1 \right) \,\left( 1 + 2\,d - \cos (4\,\theta) \right) 
  +4\,{(\cos^2(\theta) + d\, \sin^2 (\theta))}^2
     }
\eeq
which, as expected is finite. Taking the limit along the $t-$axis ($\theta=\pi/2$), we get
\beq
\Delta(t,x=0) = {\frac{1+d}{1 + 3\,d}}
= \Delta_{\rm Minkowski}
\eeq
On the other hand, taking the limit along the spatial direction:
\beq
\Delta(t=0,x) = {\frac{d\,\left( d+1 \right) }{2 + d + {d^2}}}
= \Delta_{L,{\rm Reg}}
\eeq


We also approach  this problem in another way. Since both the
point separated energy density and the correlation function
have a direction dependence, we ``average'' over the
direction. We take the hyper-spherical averaging procedure.
This involves first Wick rotating to imaginary time
$(t\rightarrow i\tau)$. Then we take the hyper-spherical
average in the Euclidean geometry and then Wick rotate back
to Minkowski space. For the energy density
\beq
\rho_E(\tau,x) =
{\frac{\Gamma\!\left({\frac{d+1}{2}}\right) }{2\, \pi^\frac{d+1}{2}}}
\frac{
    \left( d\,{{\tau}^2} - {x^2} \right)
     }{
  {{\left( {{\tau}^2} + {x^2} \right) }^{{\frac{d+3}{2}}}}\,
     }
\eeq
Now expressing $\tau=r\sin\theta$ and $x=r\cos\theta$ we do the
averaging
\beqn
\rho_E(r) &=& \frac{1}{2\pi}\int_0^{2\pi}
 \rho_E(r\sin\theta,r\cos\theta)\,d\theta
\nonumber \\
&=&
\frac{\Gamma\!\left({\frac{d+1}{2}}\right)}
       {4\, {{\pi }^{{\frac{d+3}{2}}}}\,{r^{d+1}} }
\int_0^{2\pi} \left( 
      d\,{{\sin (\theta)}^2}  -     {{\cos (\theta)}^2}
              \right)d\theta
\nonumber \\
&=& 
\frac{(d-1)\Gamma\!\left({\frac{d+1}{2}}\right)}
     {4\,\pi^\frac{d-1}{2}\,r^{d+1}}
\eeqn
W do the same for the correlation function:
\beqn
\Delta\rho^2_E(r) &=&
\frac{(d+1){\Gamma\!\left({\frac{d+1}{2}}\right)}^2}
     {32\,\pi^{d+2}\,r^{2(d+1)}}
\int_0^{2\pi} \left( d-1 + \left( d+1 \right) \,\cos (4\,\theta) \right) 
d\theta
\nonumber \\
&=&
{\frac{\left({d^2}-1 \right) \,
     {{\Gamma\!\left({\frac{d+1}{2}}\right)}^2}}
     {32\,{{\pi }^{d+1}}\, {r^{2\,\left( d+1 \right) }}}}
\eeqn
With these results, we have
\beq
\chi_{d,\rm Avg} = {\frac{1+d}{2\,\left( d-1 \right) }}
\quad{\rm and}\quad
\Delta_{d,\rm Avg} =
{\frac{1+d}{3d-1}}
\eeq
independent of whether or not we Wick rotate back to Minkowski space.
Also,
\[
\begin{array}{|c||c|c|c|c|}
\hline
d & 1 & 3 & 5 & \infty \\ \hline
\Delta_{\rm Avg} & 1 & \frac{1}{2} & \frac{3}{7} & \frac{1}{3} 
\\ \hline
\end{array}
\]
It is interesting to observe that the first set of results depend on the
direction the two points come together, and changes if one
averages over all  directions. This feature of point-separation is
known, but it could also reveal some properties of  possible  extended
structure of the underlying spacetime.


\section{Discussions}

Let us ponder on the implication of these findings pertaining to 
a) fluctuations to mean ratio and the validity of semiclassical gravity
b) the dependence of fluctuations on both the intrinsic scale 
(defined by smearing or point-separation)
and the extrinsic scale (such as the Casimir or finite temperature periodicity)
c) the treatment of divergences and meaning of regularization

\subsection {Fluctuation to Mean ratio and Validity of SCG}

 If we adopt the criterion of Kuo and Ford \cite{KuoFor} that the variance of
the fluctuation relative to the mean-squared (vev taken with respect to the ordinary
Minkowskian vacuum)  being of the order unity be an indicator of the failure of
SCG, then all spacetimes studied above would indiscriminately fall into that
category, and SCG fails wholesale, regardless of the scale these  physical
quantities are probed. This contradicts with the common expectation  that SCG
is valid at scales below  Planck energy. We believe the criterion for the validity or
failure of a theory should depend on the range or the energy probed.  Our findings here
suggest that this is indeed the case:  Both the mean (the vev of EMT) with
respect to the Minkowski vacuum) AND  the fluctuations of EMT increase as the
scale deceases.  As one probes into an increasingly finer scale or higher energy the
expectation value of EMT  grows in value and the induced  metric fluctuations
become important, leading to the failure of SCG.  A generic scale for this to
happen is the Planck length.  At such energy densities and above, particle
creation from the quantum field vacuum would become copious and their
backreaction on the background spacetime would become important \cite{scg}. Fluctuations
in the quantum field EMT entails these quantum processes. The
induced metric fluctuations \cite{stogra,ELE} renders the smooth
manifold structure of spacetime inadequate, spacetime foams \cite{Wheeler}
including topological transitions \cite{Hawking} begin to appear and SCG no
longer can provide an adequate description of these dominant processes. This
picture first conjured by Wheeler is consistent with the common notion adopted
in SCG, and we believe it is a valid one.

\subsection{Dependence of fluctuations on intrinsic and extrinsic scales}

In the previous section we have presented some detailed analysis
on the results of our calculations for the fluctuations of the energy density 
for the separate cases of Minkowski and Casimir states.  Let us now look at the bigger
picture and see if we can capture the essence of these results with some general
qualitative arguments. We want to see if  there is any simple reason behind the
following results we obtained:

a) $\Delta= O (1)$

b) the specific numeric value of $\Delta$ for the different cases

c) why  $\Delta$ for the Minkowski space from the coincidence limit of taking  a
spatial point separation is identical to the Casimir case at the coincidence
limit (6/7) and identical to the hot flat space result (2/5)  \cite{NPhD} from
taking the coincidence limit of a temporal point separation?

Point a) has also been shown by earlier calculations \cite{KuoFor,PH97}, and
our understanding is that this is true only for states of quantum nature,
including the vacuum and certain squeezed states, but probably not true for
states of a more classical nature like the  coherent state. We also emphasized
that this result should not be  used as a criterion for the validity  of
semiclassical gravity.

For point b), we can trace back the calculation of the fluctuations (second
moment) of the energy momentum tensor in ratio to its mean (first moment) to
the  integral of the term containing the inner product of  two momenta ${\bf
k}_1\cdot{\bf k}_2$ summed over all participating modes. The modes
contributing to this are different for different geometries, e.g., Minkowski
versus Casimir boundary --for the  Einstein universe this enters as 3j symbols
-- and could account for  the difference in the numerical values of $\Delta$ for
the different cases.

For point c) the difference of results between taking the coincidence limit of
a spatial versus a temporal point separation is well-known in QFTCST.  The case
of temporal split involves integration of three spatial dimensions  while the
case of spatial split involve integration of two remaining spatial  and one
temporal dimension, which would give different results. The calculation of
fluctuations involves the second moment of the field and is in this regard
similar to what enters into the calculation of moments of  inertia
\cite{Raval} for rotating objects. We suspect that the difference between the
temporal and the spatial results is similar, to the extent this analogy holds, 
to the difference in the moment of inertia of the same object but taken with 
respect to different axes of rotation. 

It may appear surprising, as we felt when we first obtained these results, that
in a Minkowski calculation the result of Casimir geometry or thermal field 
should appear, as both cases involve a scale -- the former in the spatial 
dimension and the latter in the (imaginary) temporal dimension. But if we note
that the results for Casimir geometry or thermal field are obtained at the
coincidence (ultraviolet) limit, when the scale (infrared) of the  problem does
not intercede in any major way, then the main components of  the calculations
for these two cases would be similar to the two cases of taking coincident
limit in the spatial and temporal directions in Minkowski space. All of these
cases are effectively devoid of scale as far as the pointwise field theory is
concerned. As soon as we depart from this limit the effect of the presence of a
scale shows up. The point-separated or field-smeared results for the Casimir
calculation in Sec. 4 shows clearly that the boundary scale enters in a major
way and the result for the fluctuations and the ratio are different from those
obtained at the coincident limit.  For other cases where  a scale enters
intrinsically in the problem such as   that of a massive or non-conformally
coupled field it would show a similar effect in these regards as the present
cases (of Casimir and thermal field) where a periodicity condition exists (in
the spatial and temporal directions respectively). We expect a similar strong 
disparity between point-coincident and point-separated cases. The field theory
changes its nature in a fundamental and physical way when this limit is taken.
This brings us to the second major issue brought out in this investigation,
i.e., the appearance of divergences and the  meaning of regularization in the
light of  a point-separated versus a point-defined quantum field theory.

\subsection{Regularization in the Fluctuations of EMT}

From our calculations, the smeared energy density fluctuations
for the Casimir topology has the form
\beq
\Delta\rho_L^2(\sigma) = 
   \Delta\rho_L^{\rm div} + \Delta\rho_L^{\rm cross}  + \Delta\rho_L^{\rm fin}
\eeq
with
\begin{mathletters}
\beqn
\Delta\rho_L^{\rm div}   &=& \chi_d \left( \rho_L^{\rm div} \right)^2
                        =  \chi_d (\rho(\sigma))^2
\\
\Delta\rho_L^{\rm cross} &=& 2 \chi_d \rho_L^{\rm div} \rho_L^{\rm fin}
\\
\Delta\rho_L^{\rm fin} &=& 2 \chi_{d,L} \left( \rho_L^{\rm fin} \right)^2
+{\rm terms\; that\; vanish\; as}\;\sigma\rightarrow 0
\eeqn
\end{mathletters}
where $\chi_d$ is the ratio between the fluctuations for Minkowski space
and the square of the corresponding energy density:
$\Delta\rho^2 = \chi_d(\rho(\sigma))^2$.
Our results show that
$\Delta\rho_L^2(\sigma)$ diverges as the width $\sigma$ of the smearing 
function
shrinks to zero with contributions from the truly divergent and the 
cross terms.
We also note that the divergent term 
$\Delta\rho^{\rm div}$ is state independent, in the sense that 
it is independent
of $L$, while the cross term $\Delta\rho^{\rm cross}$ is state dependent, 
as is the
finite term  $\Delta\rho^{\rm fin}$.

If we want to ask about the strength of fluctuations of the energy density, the
relevant quantity to study is the energy density correlation function
$H(x,y) = \left<\hat\rho(x)\hat\rho(y)\right>
-\left<\hat\rho(x)\right>\left<\hat\rho(y)\right>$. It is finite at
 $x\ne y$ for a linear quantum theory (this happens since the
divergences for $\left<\hat\rho(x)\hat\rho(y)\right>$ are exactly the
same as the product $\left<\hat\rho(x)\right>\left<\hat\rho(y)\right>$),
but diverges as $y\rightarrow x$,  corresponding to the
coincident or unsmeared limit $\sigma\rightarrow 0$.

To define a procedure for rendering  our expression for 
$\Delta\rho_L^2(\sigma)$ finite, one can see that there exists
choices -- which means ambiguities in the regularization scheme.
Three possibilities present themselves:
The first is to just drop the state independent 
$\Delta\rho^{\rm div}$. This is easily seen to fail since we are left with the
divergences from the cross term. The second is to neglect all terms that
diverge as $\sigma\rightarrow 0$.
This is too rash a move  since $\Delta\rho^{\rm cross}$ has, along with its divergent
parts, ones that are finite in the $\sigma\rightarrow 0$ limit. This comes about
since it is of the form $\rho_L^{\rm div} \rho_L^{\rm fin}$ and the negative
powers of $\sigma$ present in $\rho_L^{\rm div}$ will cancel out against the
positive powers in $\rho_L^{\rm fin}$.
Besides, they yield results in disagreement with earlier results using well-tested
methods such as normal ordering in flat space \cite{KuoFor} 
and zeta-function regularization in curved space \cite{PH97}.

The third choice is the one we have used in this paper. For the energy density, we can think of
regularization as computing the contribution ``above and beyond'' the Minkowski
vacuum contribution. Same for  regularizing the fluctuations.
So we need to first determine for Minkowski space vacuum
how the fluctuations of the energy density  are related
to the vacuum energy density  $\Delta\rho^2 = \chi\;(\rho)^2$.
This we obtained  for finite smearing.
For Casimir topology the sum of the divergent and cross terms take the
form
\beq
\Delta\rho_L^{\rm div} + \Delta\rho_L^{\rm cross} =
\chi\left\{
	\left(\rho_L^{\rm div}\right)^2+2\rho_L^{\rm div}\rho_L^{\rm fin}
		\right\}
= \chi\left\{
	\left(\rho_L\right)^2-\left(\rho_L^{\rm fin}\right)^2
		\right\}
\eeq
where $\chi$ is the ratio derived for Minkowski vacuum. We take this to represent
the (state dependent) vacuum contribution. What we find interesting is that 
to regularize the smeared energy density fluctuations, a state
dependent subtraction must be used. With this, just the 
$\sigma\rightarrow 0$ limit of the finite part 
$\Delta\rho_L^{\rm fin}$ is identified as the regularized fluctuations
$\Delta\rho^2_{L,{\rm Reg}}$. The ratio $\chi_L$ thus obtained gives  exactly 
the same result as derived  by Kou and Ford for $d=3$ via normal ordering 
and by ourselves for arbitrary $d$ via the $\zeta$-function.

That this procedure is the one to follow can be seen by considering the
problem from the point separation method. For this method, the energy
density expectation value is defined as the $x'\rightarrow x$ limit of
\beq
\rho(x,x') = {\cal D}_{x,x'} G(x,x')
\eeq
for the suitable Green function $G(x,x')$ and
${\cal D}_{x,x'}$ is a second order differential operator.
(For the more general stress tensor, details are reviewed in \cite{NPhD}.)
In the limit ${x'\rightarrow x}$, $G(x,x')$ is divergent. 
The Green function is regularized
by subtracting from it a Hadamard form $G^L(x,x')$:
$G_{\rm Reg}(x,x') = G(x,x') - G^L(x,x')$ \cite{Wald75}. With this, the regularized energy
density can be obtained
\beq
\rho_{\rm Reg}(x) = 
    \lim_{x'\rightarrow x}\left({\cal D}_{x,x'} G_{\rm Reg}(x,x')\right)
\eeq
Or, re-arranging terms, we can define the divergent and finite pieces as
\beq
G^{\rm div}(x,x') =  G^L(x,x'), \quad
G^{\rm fin}(x,x') = G_{\rm Reg}(x,x') = G(x,x') - G^L(x,x')
\eeq
and
\beq
\rho(x,x') = \rho^{\rm div}(x,x') + \rho^{\rm fin}(x,x')
\eeq
\[
\rho^{\rm div}(x,x')  = {\cal D}_{x,x'} G^{\rm div}(x,x')
\quad{\rm and}\quad
\rho^{\rm fin}(x,x')  = {\cal D}_{x,x'} G^{\rm fin}(x,x')
\]
so that $\rho_{\rm Reg}(x) = \lim_{x'\rightarrow x}\rho^{\rm fin}(x,x')$,
which corresponds to the $\sigma\rightarrow 0$ limit in our computation
of the Casimir energy density.

Now turning to the fluctuations, we have the point separated expression for the
correlation function
\beq
H(x,y) = \lim_{x'\rightarrow x} \lim_{y'\rightarrow y}
  {\cal D}_{x,x'} {\cal D}_{y,y'} G(x,x',y,y')
\eeq
where $G(x,x',y,y')$ is the suitable four point function. For linear theories
we use Wick's Theorem to express this in terms of products of Green functions
$G(x,x',y,y') = G(x,y) G(x',y') + {\rm permutations\; of}(x,x',y,y')$. Excluded from
the permutations is $G(x,x')G(y,y')$. (Details are in \cite{NPhD}, which includes 
correct identifications of needed permutations and Green functions.)
The general form is
\beq
H(x,y) = \lim_{x'\rightarrow x} \lim_{y'\rightarrow y}
  {\cal D}_{x,x'} {\cal D}_{y,y'} G(x,y) G(x',y')
\;+\;{\rm permutations}
\eeq
The ${(x',y')\rightarrow (x,y)}$ limits are only retained to keep track of which
derivatives act on which Green functions, but we can see there are no divergences
for $y\ne x$. However, to get the point-wise fluctuations of the energy density,
the divergences from $\lim_{y\rightarrow x}G(x,y)$ will present a problem.
Splitting the Green function into its finite and divergent pieces, we can
recognize terms leading to those we found for $\Delta\rho^2_L(\sigma)$:
\beq
H(x,y) = H^{\rm div}(x,y) + H^{\rm cross}(x,y) + H^{\rm fin}(x,y)
\eeq
where
\begin{mathletters}
\beqn
H^{\rm div}(x,y)  &=&  
    \lim_{x'\rightarrow x} \lim_{y'\rightarrow y}
    {\cal D}_{x,x'} {\cal D}_{y,y'} G^{\rm div}(x,y) G^{\rm div}(x',y')
\\
H^{\rm cross}(x,y)  &=&  
    2\lim_{x'\rightarrow x} \lim_{y'\rightarrow y}
    {\cal D}_{x,x'} {\cal D}_{y,y'} G^{\rm div}(x,y) G^{\rm fin}(x',y')
\\
H^{\rm fin}(x,y)  &=&  
    \lim_{x'\rightarrow x} \lim_{y'\rightarrow y}
    {\cal D}_{x,x'} {\cal D}_{y,y'} G^{\rm fin}(x,y) G^{\rm fin}(x',y')
\eeqn
\end{mathletters}
plus permutations.
Thus we see the origin of  both the divergent and cross terms.
When the un-regularized Green function is used,
we must get a cross term, along with the expected divergent term.
If  the fluctuations of the  energy density is 
regularized via point separation, i.e. $G(x,x')$ is replaced
by $G_{\rm Reg}(x,x')=G^{\rm fin}(x,y)$, then we should do the
same replacement for the fluctuations. When this is done, it is
only the finite part above that will be left and we can define
the point-wise fluctuations as
\beq
\Delta\rho^2_{\rm Reg}=\lim_{y\rightarrow x}H^{\rm fin}(x,y)
\eeq
The parallel with the smeared-field derivation presented here can be seen when the
analysis of $\Gz$ and $\Gperp$ in the Appendix is considered. There it is
shown they are derivatives of Green functions and can be separated into 
state-independent divergent part and state-dependent finite contribution:
 $\Gi = \Gidiv + \Gifin$, same as 
the split hereby shown for the Green function.

When analyzing the energy density fluctuations, discarding
the divergent piece is the same as subtracting from the Green function 
its divergent part. If this is done, we also no longer have the cross term, 
just as viewing the problem
from the point separation method outlined above.
We feel this makes it
problematic to analyze the cross term without also including the 
divergent term. At the same time,  regularization of the fluctuations
involving the subtraction of state dependent terms as realized in this
calculation raises new issues on regularization which merits further
investigations.

To end this discussion, we venture one philosophical point 
we find resounding  throughout all  the cases studied here.
It has to do  with the meaning of a point-defined versus
a point-separated field theory, the former we take as
an effective theory  coarse-grained from the latter, 
the point-separated theory  reflecting a finer level of
spacetime structure. 
It bears on the meaning of regularization, not
just at the level of technical procedures, but
related to finding  an effective description and
matching with physics observed at a coarser
scale or lower energy.

In particular, we feel that finding a  finite  energy momentum
tensor (and its fluctuations as we do here) which occupied
the center of attention in the research of quantum field theory 
in curved spacetime in the 70's is only a small part of a much 
larger and richer structure of theories of fields and spacetimes . 
We come to understand that whatever regularization method one 
uses to get these finite parts in a point-wise field theory should 
not be viewed as universally imparting meaning beyond its 
specified function, i.e., to identify the divergent pieces and 
provide a prescription for their removal. We believe the 
extended structure of spacetime (e.g., via point-separation or smearing )
and the field theory defined therein has its own much fuller
meaning beyond just reproducing the well-recognized result in
ordinary quantum field theory as we take the point-wise or coincident limit.
In this way of thinking, the divergence- causing terms are only
`bad' when they are forced to a point-wise limit, because of our
present inability to observe or resolve otherwise . If we accord them
with the full right of existence beyond this limit, and acknowledge
that their misbehavior is really due to our own  inability to cope,
we will be rewarded with the discovery of new physical phenomena 
and ideas of a more intricate world. (Maybe this is just another
way to appreciate the already well-heeded paths of string theory.)\\

\noindent {\bf Acknowledgement}  We thank Professor Larry Ford for interesting 
discussions, especially on the meaning of the cross term,  Dr. Alpan Raval 
for useful comments, especially on the  generic nature of the vacuum fluctuations 
to the mean, and Prof. Raphael Sorkin for discussions on the relevance of our
results to worm hole physics.
A few researchers whom we have met have voiced their doubts to us 
on the Kuo-Ford criterion, and indicated a view similar to ours as expounded here. 
We would like to thank Prof. Paul Anderson and Prof. Ted Jacobson  for 
conversations of this nature. This work is supported in part by NSF grant PHY98-00967


\appendix
\section{Evaluation of Two Point Functions for Casimir Topology}
We want to compute the smeared derivatives of the field operator
two-point functions for the Casimir geometry:
\beq
\Gperp = \left<0_L\left| \left(
        \left( \nabla_{\!\perp} \phi \right)\left(f_{\bf x}\right)
      \right)^2 \right|0_L\right>
\quad{\rm and}\quad
\Gz = \left<0_L\left| \left(
        \left( \partial_z \phi \right)\left(f_{\bf x}\right)
      \right)^2 \right|0_L\right>,
\eeq
where $\left.\left.\right|0_L\right>$ is the Casimir vacuum.
Performing the differentiation
and taking the vacuum expectation values, 
we need the integrals and sums
\begin{mathletters}
\beqn
\Gperp &=&
{\frac{l}
   {{2^d}\,{{\pi }^{{\frac{d+1}{2}}}}\,\Gamma\!\left({\frac{d-1}{2}}\right)}}
\sum_{n=-\infty}^\infty \int_0^\infty
 {\frac{{k^d}}{{\sqrt{{k^2} + {l^2}\,{n^2}}}}} {e^{-2\,\left( {k^2} + 
{l^2}\,{n^2} \right) \,{{\sigma}^2}}} \,dk
\\ \nonumber \\
\Gz &=&
{\frac{l}
   {{2^d}\,{{\pi }^{{\frac{d+1}{2}}}}\,\Gamma\!\left({\frac{d-1}{2}}\right)}}
\sum_{n=-\infty}^\infty \int_0^\infty
 {\frac{{k^{d+2}}\,{l^2}\,{n^2}}{{\sqrt{{k^2} + {l^2}\,{n^2}}}}} 
{e^{-2\,\left( {k^2} + {l^2}\,{n^2} \right) \,{{\sigma}^2}}} \,dk
\eeqn
\end{mathletters}
With the definitions of the functions
\begin{mathletters}
\beqn
\Fperp(n) &=& \int_0^\infty
 {\frac{2\,{k^d}}{{\sqrt{{k^2} + {l^2}\,{n^2}}}}} {e^{-2\,\left( {k^2} + 
{l^2}\,{n^2} \right) \,{{\sigma}^2}}} \,dk
\\ \nonumber \\
\Fz(n) &=& \int_0^\infty
 {\frac{2\,{k^{d-2}}\,{l^2}\,{n^2}}{{\sqrt{{k^2} + {l^2}\,{n^2}}}}} 
{e^{-2\,\left( {k^2} + {l^2}\,{n^2} \right) \,{{\sigma}^2}}} \,dk
\eeqn
\end{mathletters}
we use the Euler-Maclauren sum formula to re-arrange the terms 
to the more useful form ($i=x_\perp x_\perp$ or $zz$):
\beqn
\Gi &=& {\frac{l}
   {{2^d}\,{{\pi }^{{\frac{d+1}{2}}}}\,\Gamma\!\left({\frac{d-1}{2}}\right)}} 
\lim_{N\rightarrow\infty}\left(
          \frac{1}{2}\Fi(0) + \sum_{n=1}^N \Fi(n)
        \right) 
\nonumber \\
    &=& {\frac{l}
   {{2^d}\,{{\pi }^{{\frac{d+1}{2}}}}\,\Gamma\!\left({\frac{d-1}{2}}\right)}} 
\lim_{N\rightarrow\infty}\left(
            \int_0^N \Fi(n)\,dn + \frac{1}{2}\Fi(N)
           +\sum_{p=1}^q \frac{B_{2p}}{(2p)!}\left(
                       \Fi^{(2p-1)}(N)-\Fi^{(2p-1)}(0)
             \right)
        \right)
\eeqn
As we will show, $F_i(N)$
vanishes exponentially with $N$ so that  $F_i(N)$ and $F_i^{(2p-1)}(N)$ give
no contributions to the final result and we are left with
\beq
\Gi = {\frac{l}
   {{2^d}\,{{\pi }^{{\frac{d+1}{2}}}}\,\Gamma\!\left({\frac{d-1}{2}}\right)}} \left(
          \int_0^\infty \Fi(n)\,dn
        - \sum_{p=1}^q \frac{B_{2p}}{(2p)!}\Fi^{(2p-1)}(0)
      \right)
\eeq
This re-arrangement of the terms has allowed us to separate the expectation values
into terms that diverge as $\sigma\rightarrow 0$ and those that are finite in this
limit: $\Gi = \Gidiv + \Gifin$
with
\begin{mathletters}
\beqn
\Gidiv &=&   {\frac{l}
   {{2^d}\,{{\pi }^{{\frac{d+1}{2}}}}
        \,\Gamma\!\left({\frac{d-1}{2}}\right)}}\int_0^\infty F_i(n)\,dn 
\\ \nonumber \\
\Gifin &=& - {\frac{l}
   {{2^d}\,{{\pi }^{{\frac{d+1}{2}}}}
   \,\Gamma\!\left({\frac{d-1}{2}}\right)}} \sum_{p=1}^q 
\frac{B_{2p}}{(2p)!}F_i^{(2p-1)}(0)
\eeqn
\end{mathletters}

The $k$ integrations give
the explicit form of the functions $F_i(n)$:
\begin{mathletters}
\beqn
\Fperp(n) &=&
{\frac{{l^d}\,{n^d}\,\Gamma\!\left(-{\frac{d}{2}}\right)\,
     \Gamma\!\left({\frac{d+1}{2}}\right)}{{\sqrt{\pi }}}} 
{}_{1}F_{1}\left({\frac{1}{2}};1 + {\frac{d}{2}};
  -2\,{l^2}\,{n^2}\,{{\sigma}^2}\right)
\nonumber \\&&\hspace{20mm}+
{\frac{\Gamma\!\left({\frac{d}{2}}\right)}{{2^{{\frac{d}{2}}}}\,{{\sigma}^d}}} 
{}_{1}F_{1}\left({\frac{1 - d}{2}};1 - {\frac{d}{2}};
  -2\,{l^2}\,{n^2}\,{{\sigma}^2}\right)
\eeqn
and
\beqn
\Fz(n) &=&
{\frac{{l^d}\,{n^d}\,\Gamma\!\left(1 - {\frac{d}{2}}\right)\,
     \Gamma\!\left({\frac{d-1}{2}}\right)}{{\sqrt{\pi }}}} 
{}_{1}F_{1}\left({\frac{1}{2}};{\frac{d}{2}};-2\,{l^2}\,{n^2}\,{{\sigma}^2}
  \right)
\nonumber \\&&\hspace{20mm}+
\frac{l^2\,n^2}{2^{\frac{d}{2}-1}\,\sigma^{d-2}}
  \Gamma\!\left({\frac{d}{2}}-1\right) {}_{1}F_{1}\left({\frac{3}{2}} - 
{\frac{d}{2}};2 - {\frac{d}{2}};
  -2\,{l^2}\,{n^2}\,{{\sigma}^2}\right)
\eeqn
\end{mathletters}
with ${}_{1}F_{1}\left(a;b;z\right)$ the Kummer confluent hypergeometric function.

We carry out the $n$ integrations and obtain 
the divergent parts of the expectation values
\begin{mathletters}
\beqn
\Gperpdiv  &=& 
{\frac{\left( d-1 \right) \,
     \Gamma\!\left({\frac{d+1}{2}}\right)}{{2^
       {{\frac{3\,\left( d+1 \right) }{2}}}}\,d\,{{\pi }^{{\frac{d}{2}}}}\,
     \Gamma\!\left({\frac{d}{2}}\right) \, \sigma^{d+1}}}
\\
\Gzdiv  &=&
{\frac{
     \Gamma\!\left({\frac{d+1}{2}}\right)}{{2^
       {{\frac{3\,\left( d+1 \right) }{2}}}}\,d\,{{\pi }^{{\frac{d}{2}}}}\,
     \Gamma\!\left({\frac{d}{2}}\right) \, \sigma^{d+1}}}
\eeqn
\end{mathletters}

Turning to the finite contribution, we need the general form of
\beq
H =
\left. \frac{d^{2p-1}}{dn^{2p-1}} 
\left(
   A\,{n^{\beta}}
   {}_{1}F_{1}\left(\alpha;\gamma;-2\,{l^2}\,{n^2}\,{{\sigma}^2}\right)
\right)\right|_{n=0}
\eeq
To this end, we make use of the relation
\beq
\left. \frac{d^p g\left(a n^2\right)}{dn^p}  \right|_{n=0} = \left\{
 \begin{array}{ll}
  (p-1)!! \left(2a\right)^\frac{p}{2}
  g^{\left(\frac{p}{2}\right)}(0); & p\;{\rm even} \\
  0; & p\;{\rm odd}
 \end{array} \right.
\eeq
along with
\beqn
\frac{d {}_1F_1\left(\alpha,\gamma;z\right)}{dz} &=& 
   \frac{\alpha}{\gamma}{}_1F_1\left(\alpha+1,\gamma+1;z\right)
\nonumber \\
\Rightarrow
\frac{d^p {}_1F_1\left(\alpha,\gamma;z\right)}{dz^p} &=& 
   \frac{\Gamma(\alpha+p)\Gamma(\gamma)}{\Gamma(\alpha)\Gamma(\gamma+p)}
  {}_1F_1\left(\alpha+p,\gamma+p;z\right)
\eeqn
For $\gamma$ not a negative integer, ${}_1F_1\left(\alpha,\gamma;0\right)=1$.
These results lead to
\beqn
H &=&
{\frac{A\,\left(2p -1\right) !}{\Gamma\!\left(2p-\beta\right)}}
\left.\left(
    \frac{d^{2p-1-\beta} 
{}_{1}F_{1}\left(\alpha;\gamma;-2\,{l^2}\,{n^2}\,{{\sigma}^2}\right) 
}{dn^{2p-1-\beta}}
\right)\right|_{n=0}
\nonumber \\
&=&
\left\{ \begin{array}{ll}
    \frac{A\,\left(2p-1 \right) !}{\Gamma\!\left(2p-\beta\right)} 
    {2^{2p-\beta-1}}\,
    {{\left( -\left( {l^2}\,{{\sigma}^2} \right)  \right)
       }^{{\frac{2p-\beta-1}{2}}}}\,
    \left(2p-\beta-2 \right) !!
    \left.\left(
       \frac{
           d^{{\frac{2p-\beta-1}{2}} } 
           {}_{1}F_{1}\left(\alpha;\gamma;z\right) 
        }{
           dz^{ {\frac{2p-\beta-1}{2}} }
        }
    \right)\right|_{z=0};
     & \beta\quad{\rm odd} \\
    0; & \beta\quad{\rm even}
\end{array}
\right.
\eeqn
Staying with $\beta$ odd, the final result is
\beq
{\frac{A\,\left( -1 + 2\,p \right) !}{\Gamma\!\left(-\beta + 2\,p\right)}} 
{2^{2p-\beta-1}}\,{{\left( -\left( {l^2}\,{{\sigma}^2} \right)  \right)
       }^{{\frac{2p-\beta-1}{2}}}}\,\left(2p-\beta-2 \right) !!
{\frac{\Gamma\!\left(\gamma\right)\,
     \Gamma\!\left(-{\frac{1}{2}} + \alpha - {\frac{\beta}{2}} + p\right)}{
     \Gamma\!\left(\alpha\right)\,
     \Gamma\!\left(-{\frac{1}{2}} - {\frac{\beta}{2}} + \gamma + p\right)}}
\eeq
To determine the finite contributions to the smeared Green function
derivatives, we use
\[
\begin{array}{cccccc}
 && A & \beta & \alpha & \gamma \\
\Fperp&{\rm term}\;1 
                 & 
{l^d}\,\Gamma\!\left(-{\frac{d}{2}}\right)\,
     \Gamma\!\left({\frac{d+1}{2}}\right)/\sqrt{\pi }
& d
                 & {\frac{1}{2}} & 1 + {\frac{d}{2}} \\
\Fperp & {\rm term}\;2 
                 & 
{2^{{-\frac{d}{2}}}}\,{{\sigma}^{-d}}
\Gamma\!\left({\frac{d}{2}}\right)
  & 0
                 & {\frac{1 - d}{2}} & 1 - {\frac{d}{2}} \\
\Fz & {\rm term}\;1 
                 & 
{l^d}\,\Gamma\!\left(1 - {\frac{d}{2}}\right)\,
     \Gamma\!\left({\frac{d-1}{2}}\right)
/{\sqrt{\pi }}
& d
                 & {\frac{1}{2}} & {\frac{d}{2}} \\
\Fz & {\rm term}\;2 
                 & 
{2^{1 - {\frac{d}{2}}}}\,{l^2}\,{{\sigma}^{2 - d}}\,
  \Gamma\!\left(-1 + {\frac{d}{2}}\right)     
& 2
                 & {\frac{3}{2}} - {\frac{d}{2}} & 2 - {\frac{d}{2}} \\
\end{array}
\]
For $d$ odd, only the first terms of $\Fperp$ and $\Fz$ contribute. For
$d$ even, the situation involves more analysis, since for the second
terms, even though $\beta$ is even, $\gamma$ is a negative integer. This
implies ${}_1F_1\left(\alpha,\gamma,-2l^2n^2\sigma^2\right)$ divergent
structure needs to be considered as well.
For $d$ odd,
\begin{mathletters}
\beq
\Gperpfin =
- \frac{
{l^{d-1}}\,
d\,\left( d-1 \right) \,
\Gamma\!\left({-\frac{d}{2}}\right)\,\Gamma\!\left({\frac{d}{2}}\right)\,
     }{
{2^{d+3}}\,
{{\pi }^{{\frac{d+3}{2}}}}\,
     }
\sum_{p=1}
{{\left( -4l^2 \right) }^p}\,
{{\sigma}^{2\,\left(p-1 \right) }}
p\,\left(2p-1 \right) \,
{\frac{B_{2p+d-1}\,\left(2p-3 \right) !!\,
     \Gamma\!\left(p-{\frac{1}{2}}\right)}{\left( 2p+d-1\right) \,
     \left( 2p \right) !\,\Gamma\!\left(p+{\frac{d}{2}}\right)}}
\eeq
and
\beq
\Gzfin =
\frac{
	l^{d-1}\,d\,
\Gamma\!\left(-{\frac{d}{2}}\right)\,\Gamma\!\left({\frac{d}{2}}\right)\,
	 }{
	2^{d+3}\,{{\pi }^{{\frac{d+3}{2}}}} 
     }
\sum_{p=1}^\infty
{{\left( -4l^2 \right) }^p}\,
{{\sigma}^{2\,\left(p-1 \right) }}
p\,\left( 2p-1 \right) \,\left( 2p + d -2\right) \,
{\frac{B_{2p+d-1}\,\left( 2p-3 \right) !!\,
     \Gamma\!\left(p-{\frac{1}{2}}\right)}{\left( 2p+d-1\right) \,
     \left( 2\,p \right) !\,\Gamma\!\left(p+{\frac{d}{2}}\right)}}
\eeq
\end{mathletters}


\section{Plotting Smeared Casimir Results}

In this appendix we outline how $\Delta_L$ and $\Delta_{L,{\rm Reg}}$ are
manipulated so they can be plotted as a function of $\sigma$, the smearing
width.
We know $\Gi = \Gidiv + \Gifin$ and
$\rho = \rho^{\rm div} + \rho^{\rm fin}$, where
\beq
\rho^{\rm div} = {\frac{1}{32\,{{\pi }^2}\,{{\sigma}^4}}}
,\quad
\Gperpdiv = {\frac{1}{48\,{{\pi }^2}\,{{\sigma}^4}}} 
              = {\frac{2}{3}} \rho^{\rm div}
\quad{\rm and}\quad
\Gzdiv = {\frac{1}{96\,{{\pi }^2}\,{{\sigma}^4}}}
              = {\frac{1}{3}} \rho^{\rm div}
\eeq
We write $\Gi = F_i(0) + 2 \sum_{n=1}^\infty F_i(n)$ 
and $\rho_L(\sigma) = F_\rho(0) + 2 \sum_{n=1}^\infty F_\rho(n)$ 
with
\beqn
F_1 &=&
{\frac{n}{8\,{e^{8\,{n^2}\,{{\pi }^2}\,{{\sigma}^2}}}\,{{\sigma}^2}}} + 
  {\frac{{\rm Erfc}\left(2\,{\sqrt{2}}\,n\,\pi \,\sigma\right)}
    {32\,{\sqrt{2\,\pi }}\,{{\sigma}^3}}} - 
  {\frac{{n^2}\,{{\pi }^{{\frac{3}{2}}}}\,
      {\rm Erfc}\left(2\,{\sqrt{2}}\,n\,\pi \,\sigma\right)}{2\,{\sqrt{2}}\,
      \sigma}}
\nonumber \\
F_2 &=&
{\frac{{n^2}\,{{\pi }^{{\frac{3}{2}}}}\,
     {\rm Erfc}\left(2\,{\sqrt{2}}\,n\,\pi \,\sigma\right)}{2\,{\sqrt{2}}\,
     \sigma}}
\nonumber \\
F_\rho &=&
{\frac{n}{8\,{e^{8\,{n^2}\,{{\pi }^2}\,{{\sigma}^2}}}\,{{\sigma}^2}}} + 
  {\frac{{\rm Erfc}\left(2\,{\sqrt{2}}\,n\,\pi \,\sigma\right)}
    {32\,{\sqrt{2\,\pi }}\,{{\sigma}^3}}}
\eeqn
The problem here is that {\em each} of the terms in the sum over $n$ diverge
as $\sigma\rightarrow 0$.
By defining $\tilde F_i(n)= F_i(n)/X_i^{\rm div}$ 
($i=x_\perp x_\perp,zz,\rho$ and $X_i = \Gperpdiv,\Gzdiv,\rho_L^{\rm div}$) 
and $\tilde X_i = 2 {\sum'}_{n=0}^{\infty} \tilde F_i(n)-1$,
where $\sum'$ has a factor of $\frac{1}{2}$ for $n=0$,
then
\beq
X_i = X_i^{\rm div} \; \tilde X_i
\eeq
The defined functions are
\beqn
\tilde \Fperp(\sigma,n)  &=& {\frac{6\,n\,{{\pi }^2}\,{{\sigma}^2}}
    {{e^{8\,{n^2}\,{{\pi }^2}\,{{\sigma}^2}}}}} + 
  {\frac{3\,{{\pi }^{{\frac{3}{2}}}}\,\sigma\,
      {\rm Erfc}\left(2\,{\sqrt{2}}\,n\,\pi \,\sigma\right)}{2\,{\sqrt{2}}}} -
   12\,{\sqrt{2}}\,{n^2}\,{{\pi }^{{\frac{7}{2}}}}\,{{\sigma}^3}\,
   {\rm Erfc}\left(2\,{\sqrt{2}}\,n\,\pi \,\sigma\right) \nonumber \\
\tilde \Fz(\sigma,n)  &=& 24\,{\sqrt{2}}\,{n^2}\,{{\pi 
}^{{\frac{7}{2}}}}\,{{\sigma}^3}\,
  {\rm Erfc}\left(2\,{\sqrt{2}}\,n\,\pi \,\sigma\right) \nonumber \\
\tilde F_\rho(\sigma,n) &=& {\frac{4\,n\,{{\pi }^2}\,{{\sigma}^2}}
    {{e^{8\,{n^2}\,{{\pi }^2}\,{{\sigma}^2}}}}} + 
  {\frac{{{\pi }^{{\frac{3}{2}}}}\,\sigma\,
      {\rm Erfc}\left(2\,{\sqrt{2}}\,n\,\pi \,\sigma\right)}{{\sqrt{2}}}}
\eeqn
Each of these new functions to be summed over are now finite as
$\sigma\rightarrow 0$. We have divided out the known divergent part.

Now we can turn to the smeared fluctuations of the energy density. First, 
for the
sum of the divergent and cross terms we have
\beq
\Delta\rho_L^{\rm div} + \Delta\rho_L^{\rm cross} =
\frac{2}{3}
\left( ({\rho_L^{\rm div})^2} + 2\,\rho_L^{\rm div}\,\rho_L^{\rm fin}
        \right)
= \frac{2}{3} \left( 1 + 2\tilde \rho_L \right) \,\left(\rho_L^{\rm div}\right)^2
\eeq
We can clearly see how this has factored out the divergent coefficient.
For the finite term:
\beqn
\Delta\rho_L^{\rm fin} =
{\frac{({\rho_L^{\rm div})^2}}{9}} \left(
3\,\tGperp^2 + 2\,\tGperp\,\tGz + {{\tGz}^2} \right)
\eeqn
Taking together, the fluctuations can be written as
\beq
\Delta\rho_L^2 = {\frac{{(\rho_L^{\rm div})^2}}{9}} \left(
6 + 3\,{{\tGperp}^2} + 2\,\tGperp\,\tGz + {{\tGz}^2} + 
  12\,\tilde \rho_L \right)
\eeq
From this, we get the dimensionless measure
\beq
\Delta(\sigma,L) =
{\frac{6 + 3\,{{\tGperp}^2} + 2\,\tGperp\,\tGz + {{\tGz}^2} + 
     12\,\tilde \rho_L}{15 + 3\,{{\tGperp}^2} + 2\,\tGperp\,\tGz + 
     {{\tGz}^2} + 30\,\tilde \rho_L + 9\,{{\tilde \rho_L}^2}}}
\eeq
Also, considering just the finite terms
\beq
\Delta_{L,{\rm Reg}}(\sigma,L) =
    \frac{\Delta\rho_{L,{\rm Reg}}}
           {\Delta\rho_{L,{\rm Reg}} + \left(\rho_{L,{\rm Reg}}\right)^2}
=
{\frac{3\,{{\tGperp}^2} + 2\,\tGperp\,\tGz + {{\tGz}^2}}
   {3\,{{\tGperp}^2} + 2\,\tGperp\,\tGz + {{\tGz}^2} + 
     9\,{{\tilde \rho_L}^2}}}
\eeq

We can now numerically evaluate the above ratios. 
Plots of $\Delta(\sigma,L)$ and $\Delta_{L,{\rm Reg}}(\sigma,L)$ as a function
of $\sigma/L$ are presented in Figure 1. Figure 2 presents plots of
$\rho_L^{\rm fin}(\sigma,L)$ and $\sqrt{\Delta\rho_L^{\rm fin}(\sigma,L)}$.

We need to worry about the
error for $\sigma \sim 1$($=L$). Considering only the periodic $z$ direction,
this was smeared with the function
\beq
f(z,\sigma) =
\frac{1}{\sqrt{2\pi}\sigma}\exp\left(-\frac{z^2}{2\sigma^2}\right) \eeq
For an error estimate, we use $f(1,\sigma)$, for as long as this is small, then
the Gaussian smearing function does not detect the periodicity. At
$\sigma=0.4$, this error is only 4\%.


\newpage

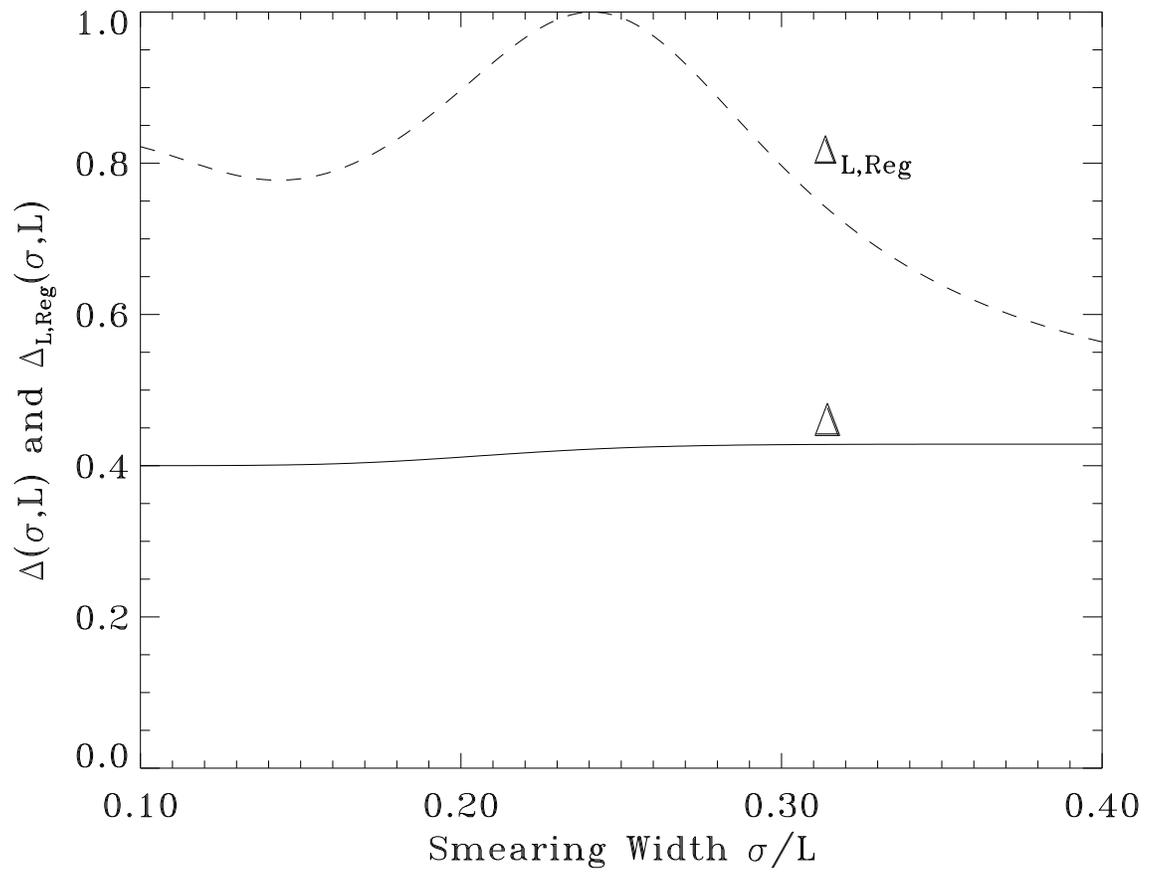
\begin{figure}
\caption{
  The dimensionless fluctuation measure
  $\Delta \equiv \left(\left<\hat\rho^2\right>-\left<\hat\rho\right>^2\right)/
                 \left<\hat\rho^2\right>$
  for the Casimir topology, along with $\Delta_{L,{\rm Reg}}$. The topology is that
  of a flat three spatial dimension manifold with one periodic dimension of period
  $L=1$. The smearing width $\sigma$ represents the sampling width of the
  energy density operator $\hat\rho(\sigma)$. $\Delta$ is for the complete fluctuations,
  including divergent and cross terms, while $\Delta_{L,{\rm Reg}}$ is just for the
  finite parts of the mean energy density and fluctuations. 
}
\end{figure}

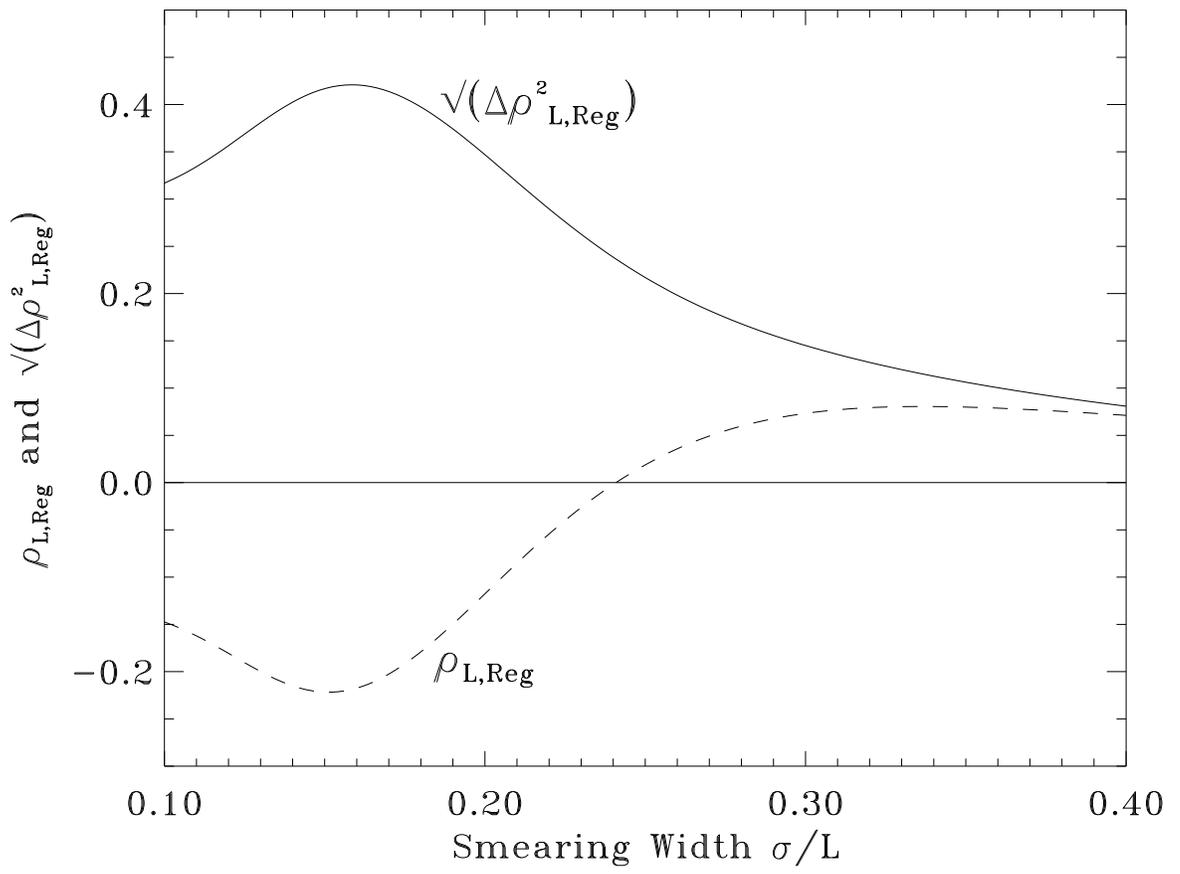
\begin{figure}
\caption{
  The finite parts of the mean energy density $\rho^{\rm fin}(\sigma,L)$ and
  the fluctuations $\Delta\rho^{\rm fin}(\sigma,L)$ for the Casimir topology, as a
  function of the smearing width. 
}
\end{figure}


\newpage
\begin{thebibliography}{999}

\bibitem{qftcst}
  N. D. Birrell and P. C. W. Davies, 
     {\sl Quantum Fields in Curved Space}
     (Cambridge University Press, Cambridge, England, 1982).

  S. A. Fulling,
      {\sl Aspects of Quantum Field Theory in Curved Space-time}
      (Cambridge University Press, Cambridge, England, 1989).

  R. M. Wald, 
      {\sl Quantum Field Theory in Curved Spacetime 
      and Black Hole Thermodynamics} 
      (The University of Chicago Press, Chicago, 1994).

  A. A. Grib, S. G. Mamayev and V. M. Mostepanenko,
      {\sl Vacuum Quantum Effects in Strong Fields} 
      (Friedmann Laboratory Publishing, St.~Petersburg, 1994).


\bibitem{Casimir} H. B. Casimir, Proc. Kon. Ned. Akad. Wet. {\bf 51},
793 (1948).

\bibitem{Barton} G. Barton, J. Phys. A. {\bf 24}, 991 (1991);
{\bf 24} 5533 (1991).

\bibitem{KuoFor}
 C.-I. Kou and L. Ford, Phys. Rev. D {\bf 47}, 4510 (1993).
  L. H. Ford, 
      Ann.\ Phys.\ {\bf 144}, 238 (1982).

\bibitem{PH97} N. G. Phillips and B. L. Hu, Phys. Rev. D. {\bf 55}, 6123
(1997).

\bibitem{Thorne}
K. S. Thorne, {\it Black Holes and Time Warps} (Norton Books, 1994)

\bibitem{HP0} B. L. Hu and   N. G. Phillips,  Int. J. Theor. Phys. {\bf 39}
 (2000) gr-qc/0004006.

\bibitem{scg} 
Ya. Zel'dovich and A. Starobinsky, Zh. Eksp. Teor. Fiz {\bf 61}, 2161 (1971)
[Sov. Phys.- JETP {\bf 34}, 1159 (1971)]
B. L. Hu  and L. Parker, Phys. Rev. {\bf D17}, 933 (1978).
F. V. Fischetti, J. B. Hartle and B. L. Hu, Phys. Rev. {\bf D20}, 1757 (1979).
J. B. Hartle and B. L. Hu, Phys. Rev. {\bf D20}, 1772 (1979).
{\bf 21}, 2756 (1980).
J. B. Hartle, Phys. Rev. D23, 2121 (1981).
P. A. Anderson, Phys. Rev. D28, 271 (1983); D29, 615 (1984).
E. Calzetta and B. L. Hu, Phys. Rev. {\bf D35}, 495 (1987)
A. Campos and E. Verdaguer, Phys. Rev. {\bf D49}, 1861 (1994).

\bibitem{valSCG}

J. B. Hartle and G. Horowitz, Phys. Rev. D 24, 257 (1981).

  R. D. Jordan,
       Phys.\ Rev.\ {\bf D36}, 3593 (1987).

 C.-I. Kou and L. Ford, Phys. Rev. D {\bf 47}, 4510 (1993).

N. G. Phillips and B. L. Hu, Phys. Rev. D. {\bf 55}, 6123 (1997).

   E. Flanagan and R. M. Wald,
          Phys. Rev. D {\bf 54}, 6233 (1996).

  W. Tichy and \'E. \'E. Flanagan,
       Phys.~Rev.~{\bf D58}, 124007 (1998).

\bibitem{stogra}

B. L. Hu, Physica {\bf A}158, 399 (1989).
B. L. Hu, Int. J. Theor. Phys.  38, 2987 (1999) gr-qc/9902064

   R. Mart\'\i n and E. Verdaguer, Int. J. Theor. Phys. 38,  3049 (1999) 
      gr-qc/9812063.  Phys.~Lett.~{\bf B465}, 113 (1999) gr-qc/9811070
        Phys.~Rev.~{\bf D60}, 084008 (1999). gr-qc/9904021. gr-qc/0001098.

  A. Roura and E. Verdaguer,
      Int.~J.~Theor.~Phys.~{\bf 38}, 3123 (1999) gr-qc/9812063

\bibitem{ELE}
    E. Calzetta and B.-L. Hu,
          Phys. Rev. D {\bf 49}, 6636 (1994).

B. L. Hu and S. Sinha,
Phys. Rev. {\bf D 51}, 1587 (1995).

B. L. Hu and A. Matacz, Phys. Rev. {\bf D51}, 1577 (1995).

A. Campos and E. Verdaguer, Phys. Rev. {\bf D53}, 1927 (1996).

    E. Calzetta, A. Campos and E. Verdaguer,
          Phys. Rev. D {\bf 56}, 2163 (1997).

  F. C. Lombardo and F. D. Mazzitelli,
       Phys.\ Rev.\ {\bf D55}, 3889 (1997).

  A. Campos and B.-L. Hu,
       Phys.~Rev.~{\bf D58}, 125021 (1998).

  A. Campos and B.-L. Hu,
      Int.~J.~Theor.~Phys.~{\bf 38}, 1253 (1999).

  E. Calzetta and E. Verdaguer,
      Phys.~Rev.~{\bf D59}, 083513 (1999).


\bibitem{states}
See any textbook in quantum optics, e.g.,
D. F. Walls and G. J. Milburn, {\it  Quantum Optics} (Springer-Verlag, Berlin, 1994).
M. O.  Scully and S. Zubairy, {\it Quantum Optics} (Cambridge University Press, Cambridge 1997).
P. Meystre and M. Sargent III, {\it Elements of Quantum Optics} (Springer-Verlag, Berlin, 1990)

\bibitem{Raval}
We thank A. Raval for this comment.

\bibitem{ptsep}

 B. S. DeWitt,  {\it Dynamical Theory of Groups and Fields} 
 (Gordon and Breach, 1965);  Phys. Rep. {\bf C19}, 295 (1975).
S. Christensen, Phys. Rev. D14, 2490  (1976); D17, 946 (1978).

S. L. Adler, J. Lieberman and Y. J. Ng, Ann. Phys. (N.Y.)  {\bf 106}, 279 (1977). 
 R. M. Wald, Phys. Rev. {\bf D17}, 1477 (1977).

\bibitem{dipolestring}
See, e.g., Zheng Yin, Phys. Lett. B466 (1999) 234. hep-th/9908152
D. Bigatti and L. Susskind,  hep-th/9908056.
N. Seiberg and E. Witten, hep-th/9908142

\bibitem{NPhD}
N. G. Phillips, Ph. D. Thesis, University of Maryland (1999)

\bibitem{Wald75}
R. M. Wald, Commun. Math. Phys. {\bf 45}, 9 (1975).

\bibitem{WuFor}
C. H. Wu and L. H.  Ford,  Phys. Rev. D60  (1999) 104013.

\bibitem{HuO'C}
B. L. Hu and D. J. O'Connor, Phys. Rev. {\bf D36}, 1701 (1987).

\bibitem{Wheeler}
J. A. Wheeler,  Ann. Phys. (N. Y.) 2, 604 (1957); 
{\it Geometrodynamics} (Academic Press, London, 1962);
in {Relativity, Groups and Topology}, eds B. DeWitt  and C. DeWitt
(Gordon and Breach, New York, 1964).

\bibitem{Hawking}
S. W. Hawking, Nucl. Phys. B144, 349 (1978).
S. W. Hawking, D. N. Page and C. N. Pope, Nucl. Phys. B170 [FS1] 283 (1980).


\end{thebibliography}
\end{document}